\documentclass{aa}

\newlength{\hfwidth}
\newlength{\hfwidthsingle}
\usepackage{graphicx}
\usepackage{color}
\usepackage{amsbsy}
\usepackage{mathrsfs}
\usepackage{mathpazo,bm}
\usepackage{amsmath}
\usepackage{multirow}

\newcommand{\aderiv}[1]{\frac{D #1}{Dt}}

\newcommand{\Ma}{\mathrm{Ma}}
\newcommand{\Rey}{\mathrm{Re}}
\newcommand{\Kn}{\mathrm{Kn}}
\newcommand{\ttimes}[1]{10^{#1}}
\newcommand{\xtimes}[2]{#1 \times 10^{#2}}

\newcommand{\vt}[1]{\mathbf{#1}}       
\renewcommand{\v}[1]{{\boldsymbol #1}} 

\newcommand{\del}{\v{\nabla}}
\newcommand{\grad}[1]{\del{#1}}
\newcommand{\Div}{\v{\nabla}\cdot}

\newcommand{\Laplace}{\nabla^2}

\newcommand{\degree}{\ensuremath{^\circ}}
\newcommand{\degreep}{\ensuremath{^\circ}~}
\renewcommand{\sun}{\odot}
\newcommand{\Eps}{{\rm Eps}}
\newcommand{\Stk}{{\rm Stk}}
\newcommand{\dv}{\Delta{\v{v}}}
\newcommand{\mearth}{\,$M_{\oplus}$}
\newcommand{\mearthp}{\,$M_{\oplus}$ }
\newcommand{\mps}{\,m\,s$^{-1}$}
\newcommand{\vrms}{$v_{\rm rms}$} 
 
\newcommand{\vesc}{$v_{\rm esc}$} 

\newcommand{\Eq}[1]{Eq. (\ref{#1})}

\newcommand{\Fig}[1]{Fig.~\ref{#1}}
\newcommand{\fig}[1]{\Fig{#1}}

\definecolor{brown}{rgb}{0.42,0.24,0.07}
\definecolor{darkgreen}{rgb}{0.0,0.6,0.00}
\definecolor{purple}{rgb}{0.7,0.0,0.7}

\begin{document}

\title{Standing on the shoulders of giants} 
\subtitle{Trojan Earths and vortex trapping\\
in low mass self-gravitating protoplanetary disks of gas and solids}

\author{
W. Lyra\inst{1},
A. Johansen\inst{2},
H. Klahr\inst{3}, \&
N. Piskunov\inst{1}
}

\offprints{wlyra@astro.uu.se}

\institute{Department of Physics and Astronomy, Uppsala Astronomical 
Observatory, Box 515, 751\,20 Uppsala, Sweden
\and Leiden Observatory, Leiden University, PO Box 9513, 2300 RA Leiden, The Netherlands
\and Max-Planck-Institut f\"ur Astronomie, K\"onigstuhl 17, 69117 
Heidelberg, Germany}

\date{Received ; Accepted}

\authorrunning{Lyra et al.}
\titlerunning{Trojan Planets}

\abstract
{Centimeter and meter-sized solid particles in protoplanetary disks are trapped
within long-lived, high-pressure regions, creating opportunities for collapse
into planetesimals and planetary embryos.}
{We aim to study the effect of the high-pressure regions generated in the
gaseous disks by a giant planet perturber. These regions consist of gas
retained in tadpole orbits around the stable Lagrangian points as a gap is
carved, and the Rossby vortices launched at the edges of the gap.}
{We performed global simulations of the dynamics of gas and solids in a low mass
non-magnetized self-gravitating thin protoplanetary disk. We employed the Pencil
code to solve the Eulerian hydro equations, tracing the solids with a large
number of Lagrangian particles, usually 100\,000. To compute the gravitational
potential of the swarm of solids, we solved the Poisson equation using
particle-mesh methods with multiple fast Fourier transforms.}
{Huge particle concentrations are seen in the Lagrangian points of the giant
planet, as well as in the vortices they induce at the edges of the carved gaps.
For 1\,cm to 10\,cm radii, gravitational collapse occurs in the Lagrangian
points in less than 200 orbits.  For 5\,cm particles, a 2\mearthp planet is
formed. For 10\,cm, the final maximum collapsed mass is around 3\mearth. The
collapse of the 1\,cm particles is indirect, following the timescale of
gas depletion from the tadpole orbits.  Vortices are excited at the edges 
of the gap, primarily trapping particles of 30\,cm radii. The rocky planet
that is formed is as massive as 17\mearth, constituting a Super-Earth. Collapse 
does not occur for 40\,cm onwards. By using multiple particle species, we
find that gas drag modifies the streamlines in the tadpole region around
the classical L4 and L5 points. As a result, particles of different radii have
their stable points shifted to different  locations. Collapse therefore takes
longer and produces planets of lower mass. Three super-Earths 
are formed in the vortices, the most massive having 4.5\mearth. 
} 
{A Jupiter-mass planet can induce the formation of other planetary embryos 
at the outer edge of its gas gap. Trojan Earth-mass planets are readily formed; although not
existing in the solar system, might be common in the exoplanetary zoo.}


\maketitle

\section{Introduction}
\label{sect:introduction}

Losing angular momentum by friction with the ambient gaseous headwind,
centimeter to meter-sized bodies in protoplanetary disks spiral
into the star on timescales as short as a hundred years (Weidenschilling 1977).
Avoiding this fate is a major unsolved problem in modern astrophysics. The
question of the formation of rocky planets is intimately connected with this
problem, since the kilometer-sized bodies (planetesimals) whence they are
believed to form (Safronov 1969) must be formed faster than the already
rapid timescale of radial drift of the rocks (0.1-1 meter-size) and boulders
(1-10 meter-size).

As colliding boulders have very poor sticking properties (Benz 2000), a
possible scenario for the formation of planetesimals is direct gravitational
collapse of the layer of boulders (Goldreich \& Ward 1973). This
hypothesis has met with criticism because no route for achieving critical
densities could be found (Weidenschilling \& Cuzzi 1993), but it has
recently gained momentum due to a series of advances in modeling
the coupled dynamics of gas and boulders through both analytical
calculations and numerical simulations. Youdin \& Goodman (2005) showed that
when rocks and boulders migrate due to the drag force, they trigger a streaming
instability that develops into a traffic jam in their migrating flow, with
dramatic effects for the particle concentrations (Johansen et al.\
2006b, Paardekooper 2006, Johansen \& Youdin 2007, Balsara et al.\
2008). Fromang \& Nelson (2005) modeled the dynamics
of particles in magnetized global disks and showed that trapping occurs in the
pressure maxima of the turbulence generated by the magneto-rotational
instability (MRI). The number of particles, however, was too low ($\leq$
3000) to say anything about possible gravitational collapse. Johansen et al.\
(2006a) simulated the flow in an MRI-active local box using a statistically
significant number of particles ($\ttimes{6}$), and showed that the particle
concentration is high enough to achieve critical densities. 

Studies with self gravity to follow the collapse are restricted to local boxes
(Johansen et al.\ 2007) and the massive disk case (Rice et al.\ 2004,
Rice et al.\ 2006).  The former couples the effects of particle concentrations
due to the streaming instabilities with those due to the turbulence generated
by the MRI to show that the turbulent layer of boulders locally collapse into
dwarf planets on very short timescales. The latter is a global disk calculation
of marginally gravitationally unstable gaseous disks, where boulders are shown
to concentrate in the spiral arms that develop, where they also achieve
critical density. 

A broad conclusion that can be drawn from these studies is that any region
with higher pressure than its surroundings tends to concentrate solids 
(Haghighipour \& Boss 2003). Therefore, in order to trigger collapse of 
the solids, one has to create long-lived, high-pressure regions in the 
gas phase. A perturber, then, is expected to have major consequences in 
the dynamics of embedded rocks and 
boulders. Paardekooper \& Mellema (2004) studied the dynamics of dust in a
gaseous disk, finding that even low mass planets carve a deep dust gap. An
update by Paardekooper (2007) showed an interesting feature. As early as 20
orbits, meter-sized particles tend to concentrate at the gap edges and at
co-rotation in tadpole orbits. However, as Lagrangian trapping was not the main
scope of their study, they did not further assess the consequences of particle
concentration in 1:1 resonance, focusing instead on the other mean motion
resonances brought about by the planet. 

Fouchet et al.\ (2007) also explored the same problem, in 3D SPH simulations,
considering not only different particle radii, but also different masses for
the perturber. The results are very similar to those of Paardekooper (2007),
but they argue against the accumulation they see being the result of resonance
trapping.  They come to this conclusion because the signatures of resonance
trapping, easily identifiable in a eccentricity vs. semi-major axis plot for
decoupled particles, disappears when one considers gas drag. Instead, they
claim that it occurs more likely due to the dust concentrating at the gas
pressure maxima at the edges of the gap. Fouchet et al.\ (2007) also notice
that the 1\,m sized boulders are found in 1:1 resonance at later times. They
speculate that the same occurs for other particle sizes they considered (10\,cm
and 1\,cm sized pebbles), but as the dust gap in this case was too narrow
compared to the extended disk they considered (20\,AU), they could not spot the
rocks trapped in the co-orbital region. 

One possibility that was unexplored in these works is whether a direct
collapse can occur at the enhanced particle concentrations. There are
significant gas overdensities in co-rotation, especially at the Lagrangian
points, for at least 200 orbits (de Val-Borro et al.\ 2006). In these regions,
the solids are subject to drag forces for a period long enough to allow
concentration and eventual collapse into kilometer-sized bodies in 1:1
resonance. In this paper, we show that the trapping provided in the Lagrangian
points is so efficient that the final mass of the collapsed body is that of
terrestrial planets. 

The collapse of the solids that get trapped at the edges of the gas gap is also
an interesting issue. As shown by de Val-Borro et al.\ (2007), the gap that the
planet carves has a density gradient propitious to the excitation of the Rossby
wave instability (RWI; Li et al.\ 2001). The anti-cyclonic vortices that form
are entities of great interest, since they induce a net force on solid
particles toward their centers, raising the local solids-to-gas ratio and 
favoring gravitational collapse (e.g. Barge \& Sommeria 1995, 
Bracco et al. 1999, Chavanis 2000). We show in this paper that the combination of the 
particle concentration seen by Paardekooper (2007) and Fouchet et al.\ (2007), 
together with the vortices predicted by de Val-Borro et al.\ (2007) lead to a 
powerful particle trap, raising the density of solids
by three to four orders of magnitude. The collapse leads not to a 
kilometer-sized body or to a dwarf planet, but to masses comparable to that of the
terrestrial planets and in some cases, super-Earths. 

An initial step towards modeling this scenario was put forth by Beaug\'{e} et
al.\ (2007). In this recent study, these authors perform pure $N$-body
calculations of a few number (usually 500) of dwarf planets of 0.3 $M_{\rm
Moon}$ around the L4 point of Jupiter. They find that a reasonable fraction of
the bodies escape the tadpole orbit by close encounters with the giant. The
rest of the particles successfully concentrate into a single Trojan planet, but
no more massive than 0.6\mearth. They do not solve concurrently for the
dynamics of gas and solids, but they assess how the formation process would
work in a gas rich scenario by performing planet-disk simulations and verifying
the gas conditions around the Lagrangian points. The densities and velocities
are then used to quantify coefficients for the drag laws.  This {\it ad hoc}
drag force is then added in the pure $N$-body calculations. 

In this paper we model gas and dust self-consistently, using
$\ttimes{5}$ particles to
represent the solids phase. Unlike Beaug\'{e} et al.\ (2007), we do not assume
the particles to be as massive as dwarf planets. Instead, we treat them as
meter-sized bodies, their gravitational potential computed by solving
the continuous Poisson equation. Although the formation of Trojans is
our primary interest, we model a radially extended region of the disk,
and are able to explore the gap edge as well, where the
anti-cyclonic vortices form.

In the next sections we describe the model equations, the Poisson solver and
drag law used. The results are presented in 
Sections~\ref{sect:single-species}-\ref{sect:neptune}, followed by
a concluding discussion in Sect.~\ref{sect:conclusions}.

\section{Dynamical equations}
\label{sect:dynamical-equations}

\begin{figure*}
\begin{center}
  \resizebox{\textwidth}{!}{\includegraphics{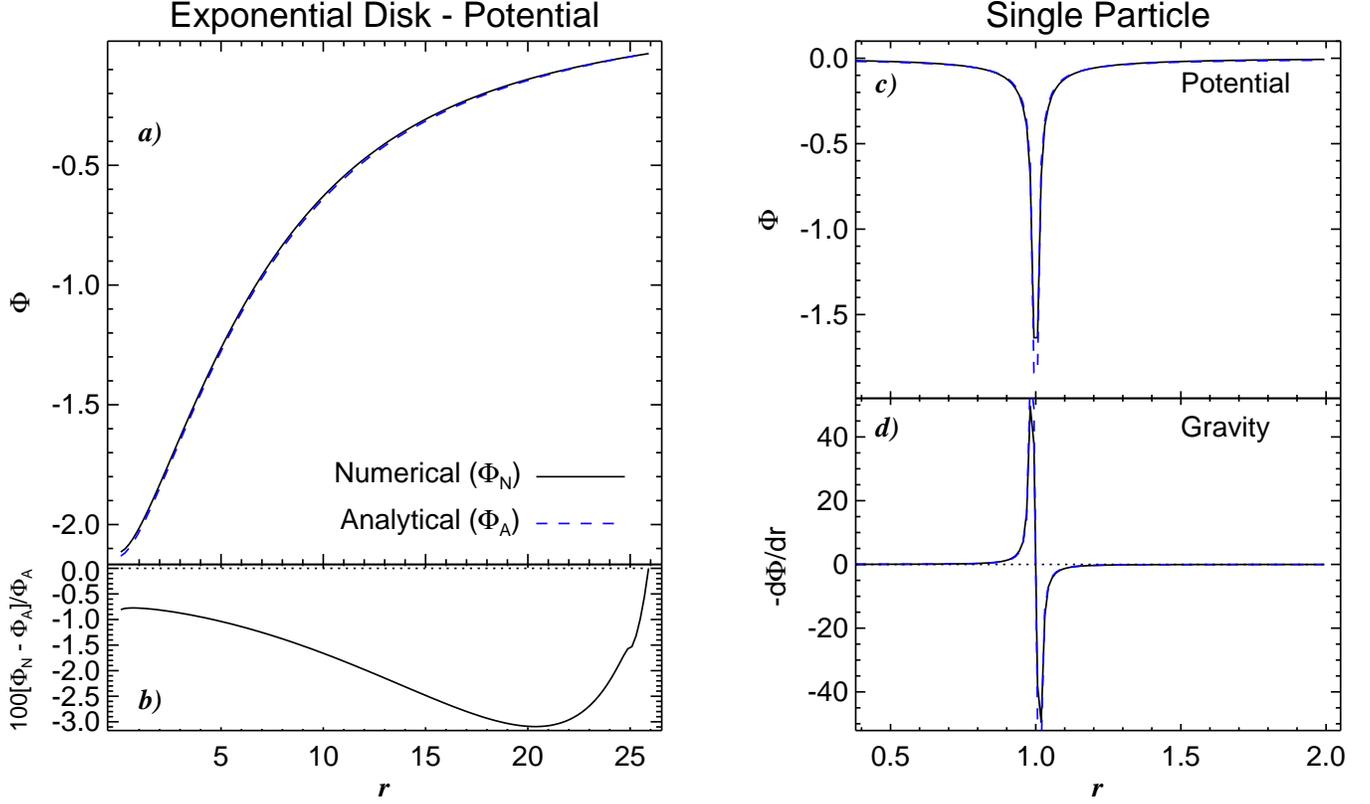}}
\end{center}
\caption[]{$a$. The potential generated by an exponential disk computed by
\Eq{eq:poisson-fourier} is compared with the analytical expression. The
accuracy ($b$.) is at the percent level. \\$c$. The potential generated by a
single particle agrees very well with its Newtonian prediction.  In particular,
the scheme ensures that the gravity ($d.$) is smooth and the particle does not
suffer self-acceleration.}
\label{fig:potential} 
\end{figure*}

We work in the thin disk approximation, using the vertically averaged equations
of hydrodynamics

\begin{eqnarray}
\aderiv{\varSigma_g} &=& -\varSigma_g{\Div\v{u}} \ + \  f_{_D}(\varSigma_g)    \label{eq:continuity}\\
\aderiv{\v{u}} &=& \ -\frac{1}{\varSigma_g}\del{P} \ - \ \del\varPhi \ - \ \frac{\varSigma_p}{\varSigma_g}\v{f}_d \ + \ \v{f}_{\nu}(\v{u},\varSigma_g)  \label{eq:Navier-Stokes}\\
\frac{d{\v{v}}_{p}}{dt} &=& - \del\varPhi + \v{f}_d                                             \label{eq:particle}\\
\frac{d{\v{x}}_{p}}{dt} &=&  \v{v}_{p}                                            \label{eq:particle-pos}\\
\varPhi&=&\varPhi_{\rm sg} -\sum_{i}^{n}{\frac{GM_i}{\sqrt{{\mathcal R}_i^2+b_i^2}}}               \label{eq:potential}\\
\Laplace\varPhi_{\rm sg} &=& 4{\pi}G\varSigma\delta(z)    \label{eq:poisson}\\        
P&=&\varSigma_g c_s^2                                                                   \label{eq:state}\\
f_d &=& - \left(\frac{3\rho_g C_D |\dv|}{8a_\bullet \rho_\bullet}\right)\dv.   \label{eq:drag-acelleration}
\end{eqnarray}

In the above equations, $\varSigma_g$ and $\varSigma_p$ are the vertically
integrated gas density and bulk density of solids, respectively. In
\Eq{eq:poisson}, $\varSigma$ is their sum. $\v{u}$ stands for the velocity of
the gas parcels; $\v{v}_p$ is the velocity of the solid particles, and
$\v{x}_p$ is their position; $P$ is the vertically integrated pressure,
$c_s$ is the sound speed, $\varPhi$ the gravitational potential and $\v{f}_d$
is the drag force by which gas and solids interact.  In
\Eq{eq:drag-acelleration}, $\rho_\bullet$ is the internal density of a solid
particle, $a_\bullet$ its radius, and $\dv=\v{v}_p-\v{u}$ its velocity relative
to the gas. The nature of the drag is concealed in the dimensionless
coefficient $C_D$, discussed in section~\ref{sect:drag-force}. The operator ${D}/Dt = \partial
/{\partial}t + \v{u}\cdot\del$ represents the advective derivative. 

The gravitational potential $\varPhi$ has contributions from the star, the
giant planets, and the disk's self-gravity. The star and the planets are
treated as massive particles with a simple $N$-body code. In \Eq{eq:potential},
$G$ is the gravitational constant, $M_i$ is the mass of particle $i$ and
${\mathcal R}_i=|\v{r}-{\v{r}}_{p_i}|$ is the distance relative to particle
$i$. The quantity $b_i$ is the distance over which the gravity field of
particle $i$ is softened to prevent singularities.

The function $f_{_D}(\varSigma_g)$=$D_3 \nabla^6 \varSigma_g$ is a third
order hyperdiffusion term. In Fourier space, it is proportional to
$k^6$, where $k$ is the wavenumber. Being so, it behaves as a
high-frequency filter, and is very effective in providing numerical
stabilization near the grid scale while having little effect in the more
quiescent larger scales. The function $\v{f}_{\nu}(\v{u},\varSigma_g)$ has both
a hyperviscosity and a shock viscosity term

\begin{eqnarray}
  \v{f}_{\nu}(\v{u},\varSigma_g)&=&\varSigma_g^{-1} \Div{\left(2\varSigma_g\nu_3\vt{S}^{(3)}\right)} + \nonumber\\
  & & \zeta_\nu \left[\grad{(\Div{\v{u}})}+ (\grad{\ln \varSigma_g} + \grad{\ln \zeta_\nu})\Div{\v{u}}\right] \label{eq:shockvisc}
\end{eqnarray}where $S^{(3)}_{ij}$=$\partial_j^5 u_i$ is a simplified (third-order)
rate-of-strain tensor and the shock term $\zeta_\nu$ follows the formulation of
Haugen et al.\ (2004), being proportional to the smoothed (over three grid cells
in each direction) maximum (also over three grid cells) of the positive part of
the negative divergence of the velocity, i.e.
\begin{equation}
  \zeta_{\nu}=\nu_{\rm shock} \left<\max_3[(-\Div\v{u})_+]\right>\left(\Delta x\right)^2.
\end{equation}
The shock viscosity coefficient $\nu_{\rm shock}$ is a parameter of order
unity. We use $\nu_{\rm shock}$=1 and $\nu_3$=$D_3$=$\xtimes{5}{-12}$. This
hyperviscosity relates to the usual Laplacian viscosity by $\nu_3$=$\nu k^4$.
Therefore, it corresponds to $\nu\simeq\xtimes{3}{-3}$ (or $\alpha\simeq 1$) at
the grid scale where $k$=$\pi/\Delta{x}$, and $\nu\simeq \ttimes{-11}$
($\alpha\simeq\xtimes{4}{-9}$) at the largest scale of the box. Here
$\alpha$=$\nu\varOmega_K c_s^{-2}$ is the usual Shakura-Sunyaev viscosity
parameter (Shakura \& Sunyaev 1973) and $\varOmega_K$ the Keplerian frequency. 

The simulations were done with the Pencil Code{\footnote{See
http://www.nordita.org/software/pencil-code}} in Cartesian and
cylindrical geometry. We write Cartesian coordinates as ($x$,$y$) and
cylindrical coordinates as ($r$,$\phi$). 

\subsection{Self-gravity}
\label{sect:self-gravity}

We solve the Poisson equation \Eq{eq:poisson} using the traditional rapid
elliptic solvers with multiple Fast Fourier Transforms.
For a single Fourier component
$\widehat\varSigma$ the solution to \Eq{eq:poisson} is
\begin{equation}
  \widehat\varPhi = -\frac{2\pi{G}\widehat\varSigma}{|\v{k}|},
  \label{eq:poisson-fourier}
\end{equation}
where $\v{k}=(k_x,k_y)$ is the in-plane wavenumber and the hat denotes
Fourier transformed quantities. The potential is then found by taking the
inverse transform to real space.

As the Fourier transform assumes periodic boundaries, the potential derived is
as if the disk was accompanied by mirror images of itself, the gravity of these
images influencing the motion of the fluid. To reduce this problem, we
expand the grid by a factor 2 prior to solving the Poisson equation. In this
expanded grid, the mirrors are still present, but they are now located
so far away from the regions of interest that no spurious behavior is
introduced by the periodic boundaries. We show in \Fig{fig:potential}a the
potential of an exponential disk, typical of galaxies, in which case the
analytical solution is well known (Freeman 1970). The deviations are at the
percent level, as seen in \Fig{fig:potential}b. 

The gravitational potential of the swarm of particles is found by the same
method outlined above. The surface density of particles is assigned to
the mesh  using the Triangular Shaped Cloud (TSC) scheme (Hockney \& Eastwood
1981, Youdin \& Johansen 2007), whereby the influence of a particle is
assigned to three grid points in each direction.  After finding the potential,
the acceleration is interpolated back to the position of the particles, using
the same TSC scheme, to avoid self-acceleration (Johansen et al.\ 2007).

Analytical prediction and numerical solution for the potential of a single
particle are compared in \Fig{fig:potential}c. Deviations occur only near the
particle position, as expected for a particle-mesh method. \Fig{fig:potential}d
shows the gravitational acceleration generated by this potential. The
agreement is excellent and the particle does not experience any
self-acceleration. 

\subsection{Drag force}
\label{sect:drag-force}

Solid particles and gas exchange momentum due to interactions that
happen at
the surface of the solid body. The many processes that can occur are generally
described by the collective name of ``drag'' or ``friction''. The drag regimes
are controlled by the mean free path $\lambda$ of the gas, which can be
expressed in terms of the Knudsen number of the flow past the particle
$\Kn=\lambda/(2a_\bullet)$. High Knudsen numbers correspond to free molecular
flow, or Epstein regime. Stokes drag applies at low Knudsen numbers. In
this section we describe our numerical implementation of drag forces in the
Pencil Code for general values of $\Kn$. We use the formula of Woitke \& Helling
(2003; see also Paardekooper 2007), which interpolates between Epstein and
Stokes regimes
\begin{equation}
  C_D = \frac{9\Kn^2 C_D^\Eps + C_D^\Stk}{(3\Kn+1)^2}.
  \label{eq:coeff-general}
\end{equation}
where $C_D^\Eps$ and $C_D^\Stk$ are the coefficients of Epstein and Stokes
drag, respectively. They read 
\begin{eqnarray}
  C_D^\Eps &\approx& 2\left(1+\frac{128}{9\pi \Ma^2}\right)^{1/2} \label{eq:coeff-epstein}\\
  C_D^\Stk&=&\left\{ \begin{array}{ll}
    24\,\Rey^{-1} + 3.6\,\Rey^{-0.313}  & \mbox{; $\Rey \leq 500$};\\
    \xtimes{9.5}{-5}\,\Rey^{1.397} & \mbox{; $500 < \Rey \leq 1500$};\\
    2.61  & \mbox{; $\Rey > 1500$}. \end{array} \right. \label{eq:coeff-stokes}
\end{eqnarray}
where $\Ma=|\dv|/c_s$ is the Mach number, $\Rey= 2a_\bullet\rho_g |\dv|/\mu$ is
the Reynolds number of the flow past the particle, and $\mu=\sqrt{8/\pi}\rho_g
c_s \lambda/3$ is the kinematic viscosity of the gas.

The approximation for Epstein drag (Kwok 1975) connects regimes of low and high
Mach number ($\Ma=|\dv|/c_s$) to good accuracy, and is more numerically
friendly than the general case (Baines et al.\ 1965). The piece-wise function
for the Stokes regime are empirical corrections to Stokes law
($C_D=24\Rey^{-1}$), which only applies for low Reynolds numbers.

\begin{figure*}
  \begin{center}
    \resizebox{\hfwidth}{!}{\includegraphics{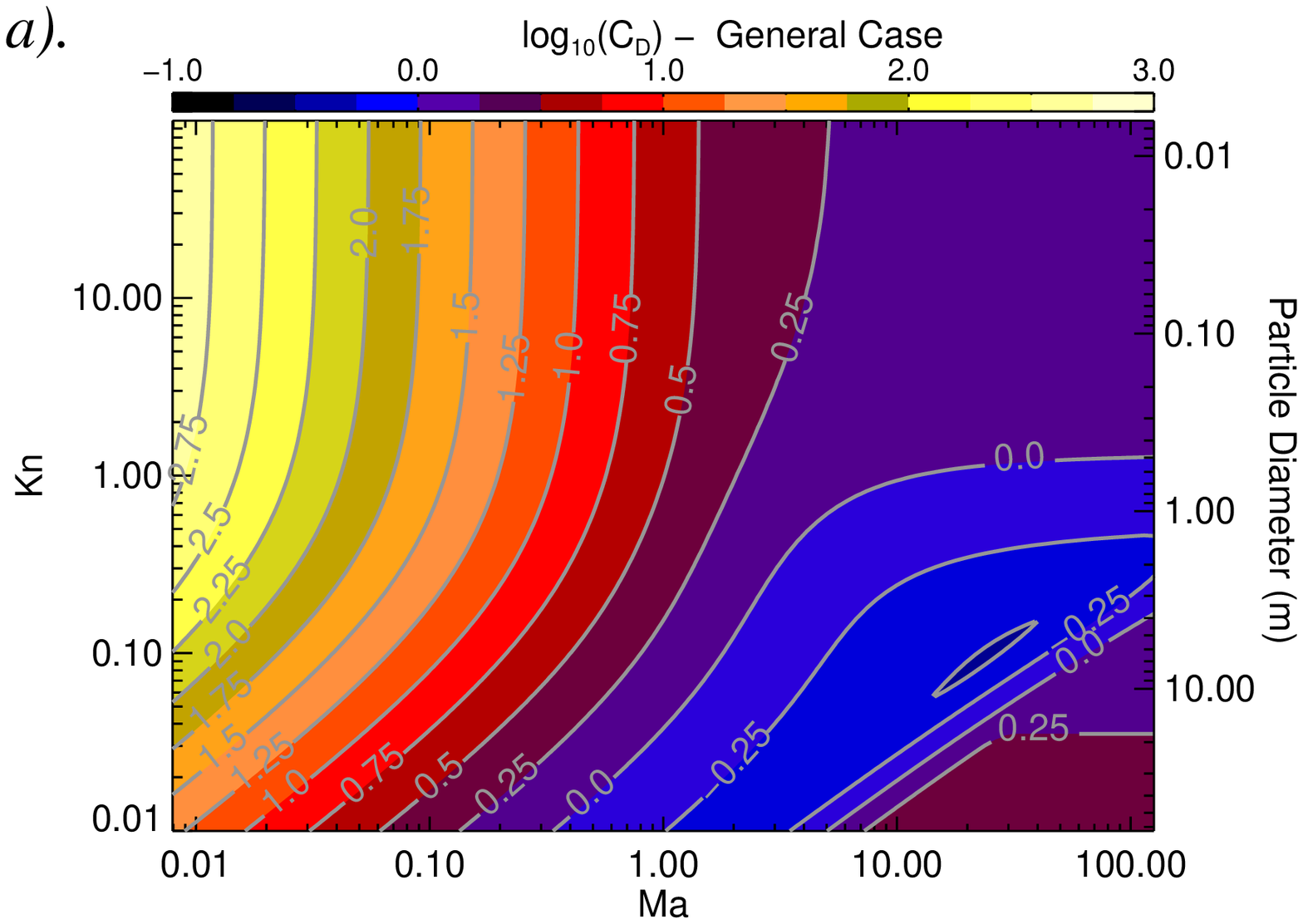}}
    \resizebox{\hfwidth}{!}{\includegraphics{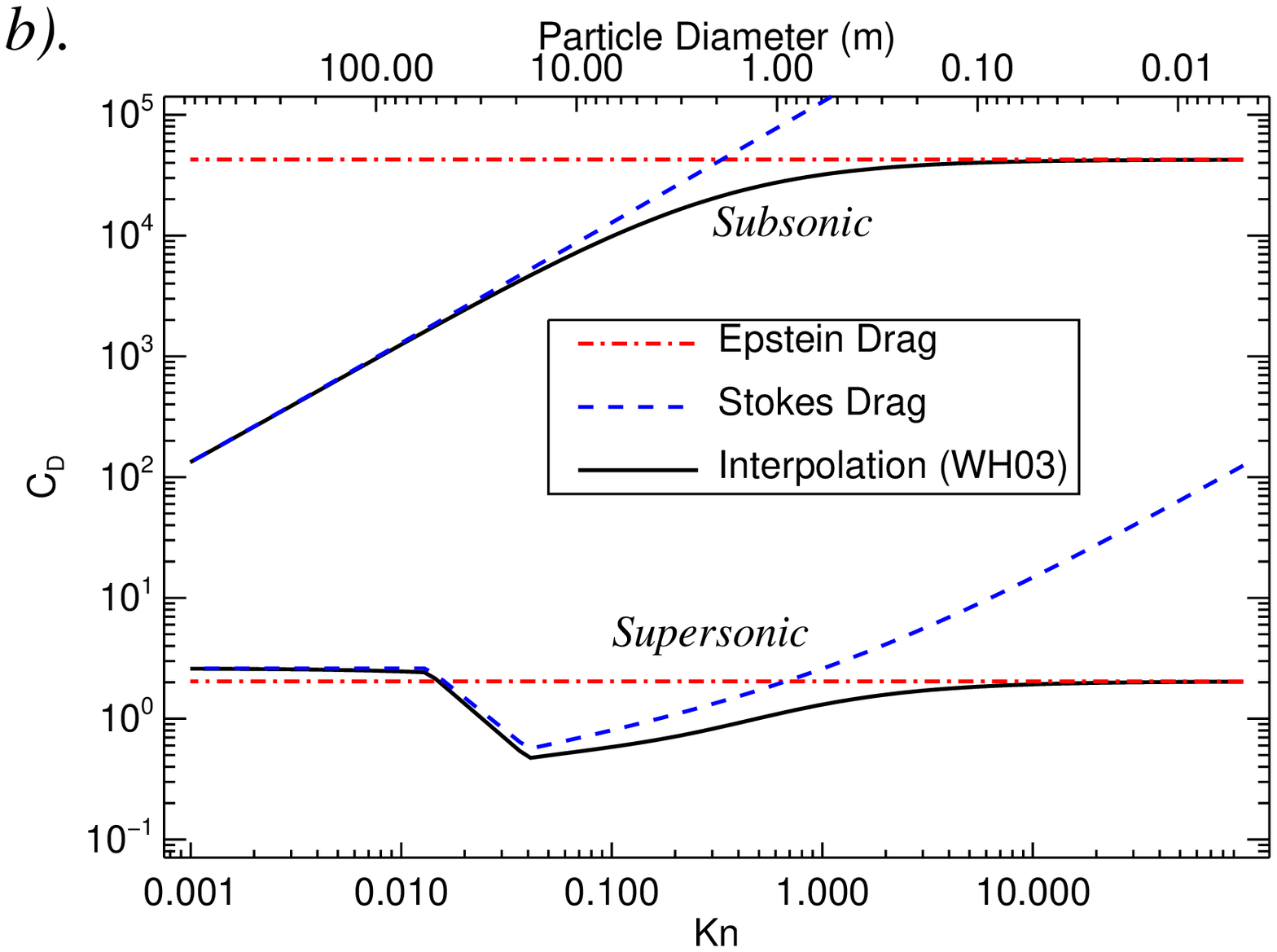}}
  \end{center}
  \caption[]{The interpolated general drag coefficients in {\it a).} the plane
  of Knudsen and Mach numbers. In {\it b).} we show two slices at
  subsonic and supersonic motion, comparing to the respective predictions of
  Epstein and Stokes drag. For particles up to 10 centimeters, Epstein drag
  does not deviate much from the general (interpolated) coefficient. Pure
  Stokes drag starts to apply only beyond 10 meters.}
  \label{fig:frictiontime}
\end{figure*}

\Fig{fig:frictiontime}a shows the value of this coefficient in the plane of
Mach and Knudsen numbers.  As stressed by Woitke \& Helling (2003), at
intermediate Knudsen numbers, the true friction force yields smaller values
than in both limiting cases, which is illustrated in \Fig{fig:frictiontime}b.
Another measurement of the strength of the drag force is the friction
time $\tau_{_f}$, defined as the inverse of the quantity in parentheses in
\Eq{eq:drag-acelleration}
\begin{equation}
  \tau_{_f}=\frac{4 \lambda \rho_\bullet}{3 \rho_g C_D c_S}\frac{1}{\Ma \Kn} .
  \label{eq:friction-time}
\end{equation}
The drag acceleration can then be cast in the compact form 
\begin{equation}
  f_d=-\frac{1}{\tau_{_f}}\Delta{\v{v}_p}.
  \label{eq:drag}
\end{equation}

\begin{figure*}
  \begin{center}
    \resizebox{\hfwidth}{!}{\includegraphics{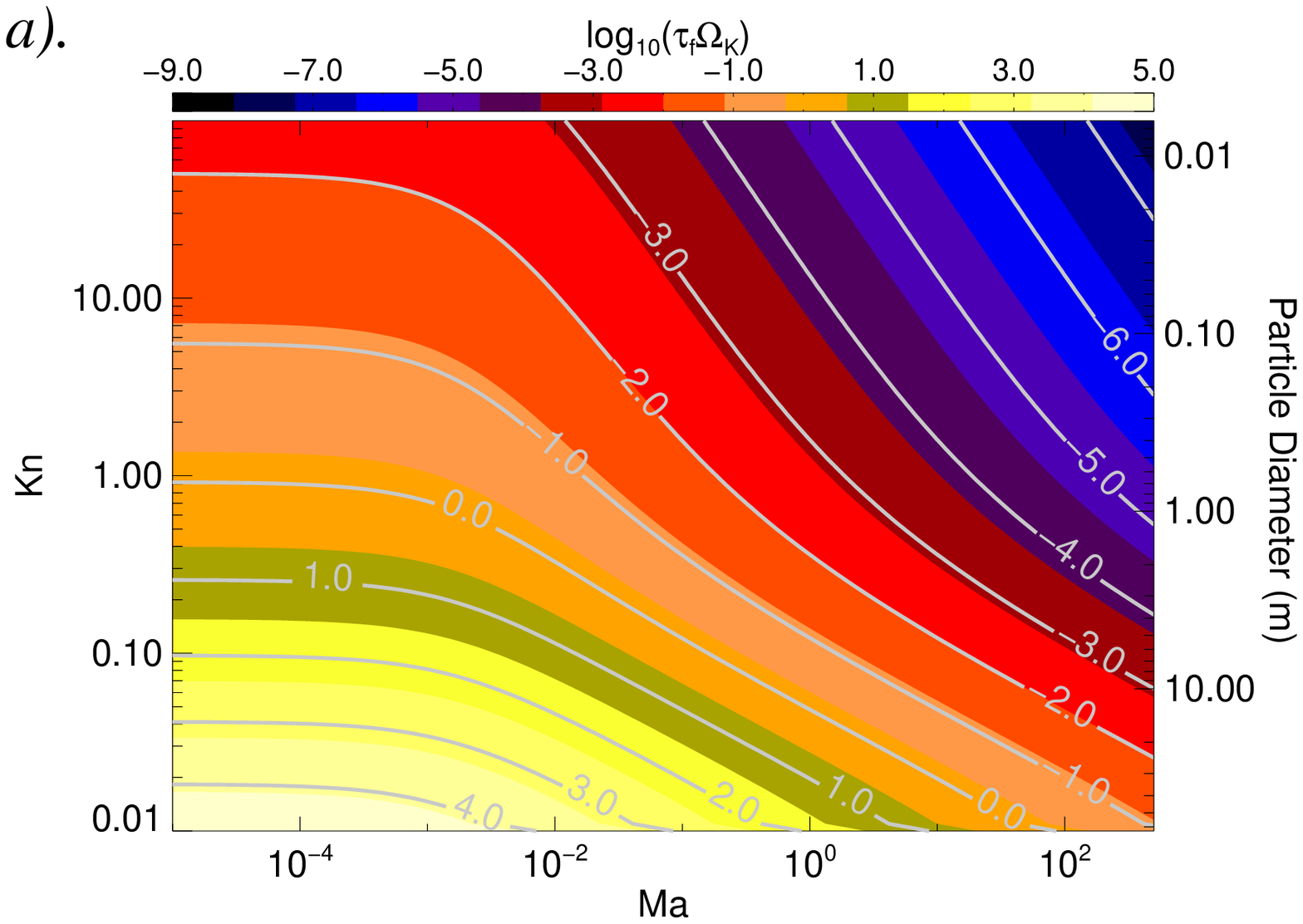}}
    \resizebox{\hfwidth}{!}{\includegraphics{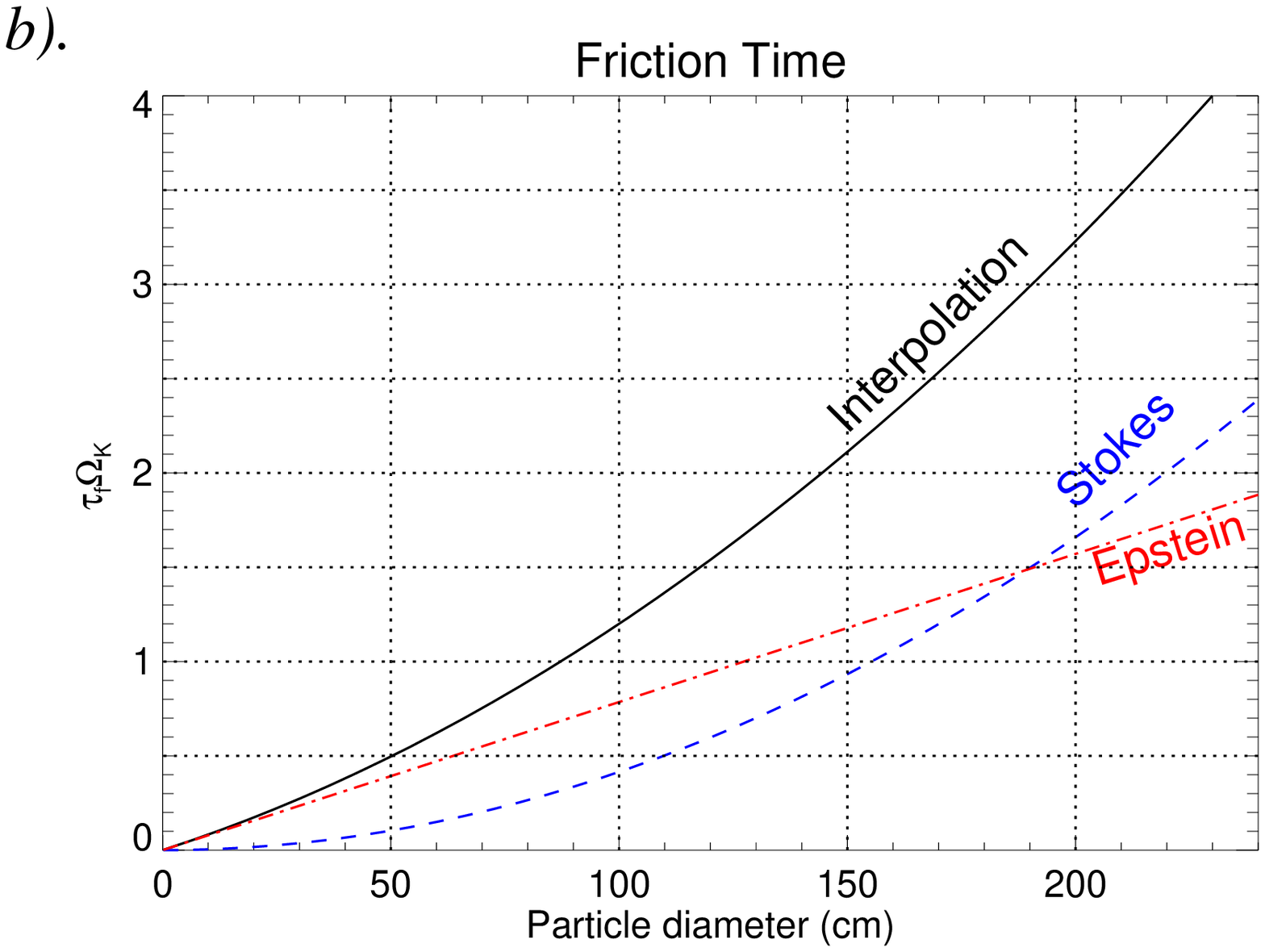}}
  \end{center}
  \caption[]{{\it a).} The drag coefficient translated to friction
  times for our choice of numerical parameters at Jupiter's orbital position.
  In {\it b).} we show a slice of {\it a).} at
  subsonic regime, where the predictions of Epstein and Stokes law are shown
  for comparison. The general drag yields more loose coupling than both
  limiting cases.}
  \label{fig:frictiontime2}
\end{figure*}

\begin{figure*}
  \begin{center}
    \resizebox{.95\textwidth}{!}{\includegraphics{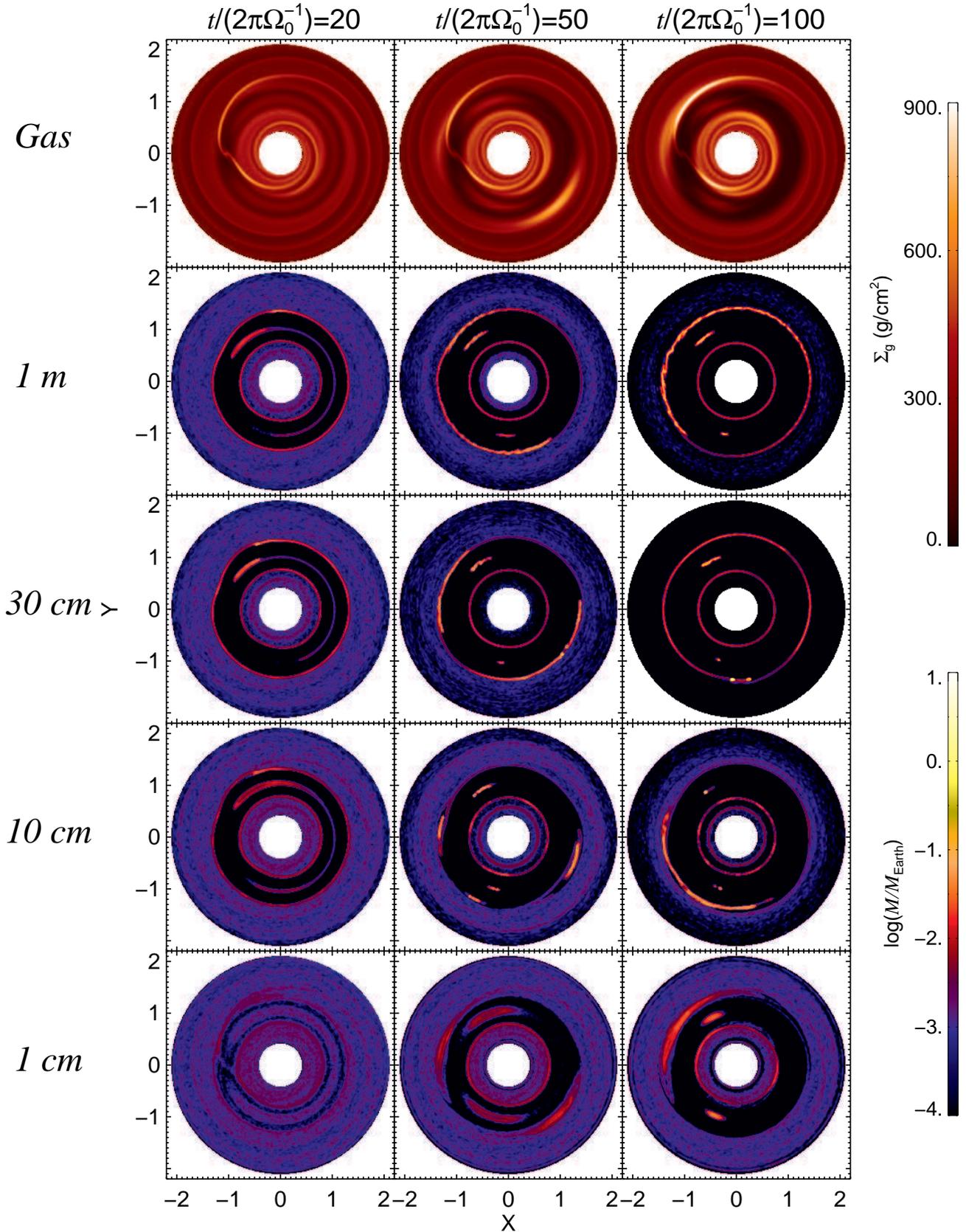}}
  \end{center}
  \caption[]{Snapshots of the gas and solid phase of the disk for several
  single particle species runs, for a perturber of Jupiter's mass. At the end of
  the simulations at 100 orbits, the swarms of particles in the L4 and L5
  points of the $a_\bullet$=1\,m run remain unbound. The 10\,cm and 30\,cm
  particles underwent collapse at the Lagrangian points, with the fragmentation
  being more efficient for the 10\,cm particles than for the 30\,cm ones. In
  the $a_\bullet$=10\,cm case, the particles underwent collapse in both
  Lagrangian points, L5 harboring a 2.6\mearthp planet, L4 a 0.6\mearth. At the
  edges of the gap, even $a_\bullet$=1\,cm particles are trapped within the
  vortices. In the $a_\bullet$=10\,cm run, the effect of the anti-cyclonic
  motion lead to a final collapsed mass of 0.3\mearth.  When the vortices merge
  into a single giant vortex, the $a_\bullet$=30\,cm particles are seen to have
  undergone runaway growth of solids, reaching 17\mearth.}
  \label{fig:jupiter}
\end{figure*}

\begin{figure*}
  \begin{center}
    \resizebox{\hfwidth}{!}{\includegraphics{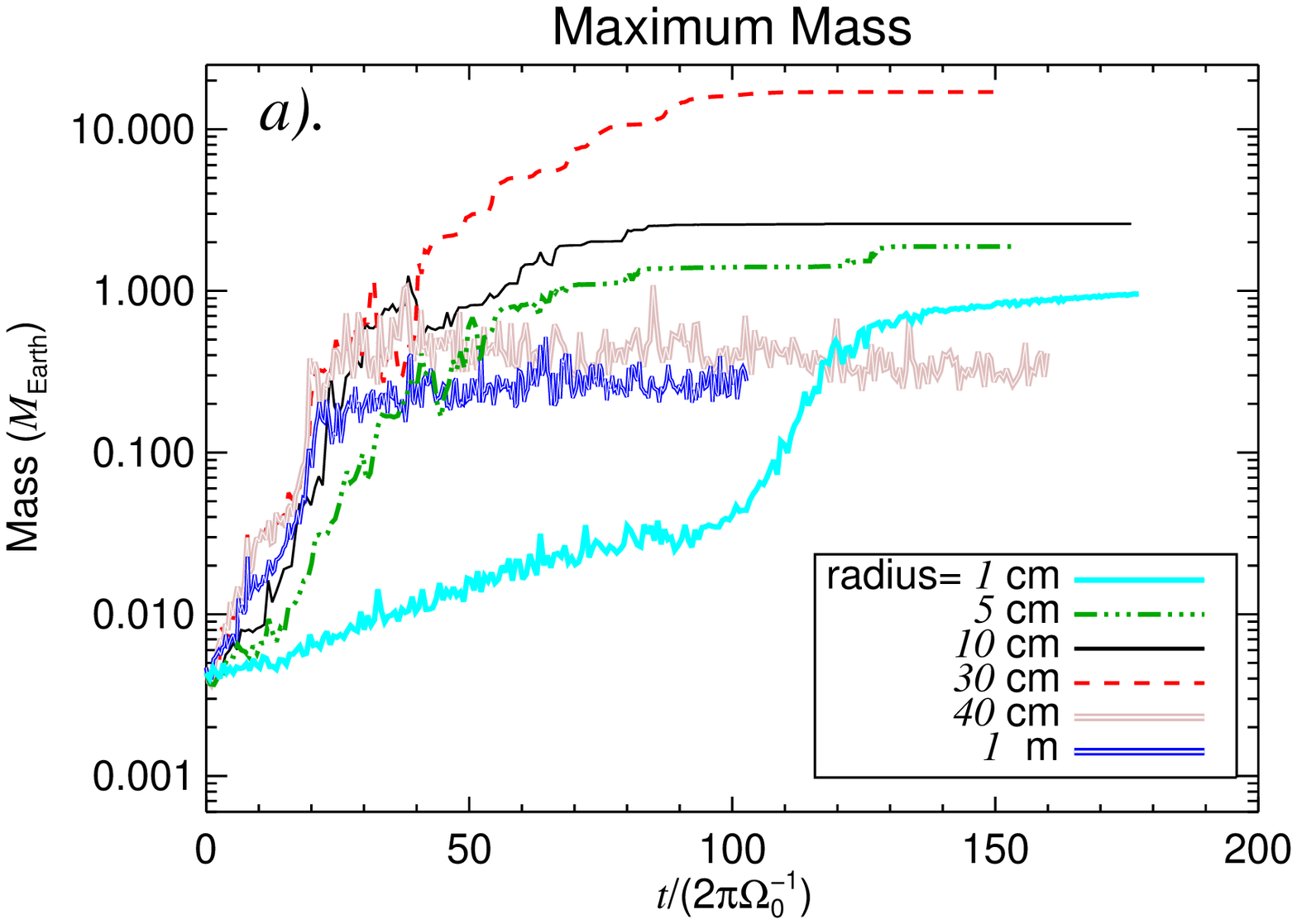}}
    \resizebox{\hfwidth}{!}{\includegraphics{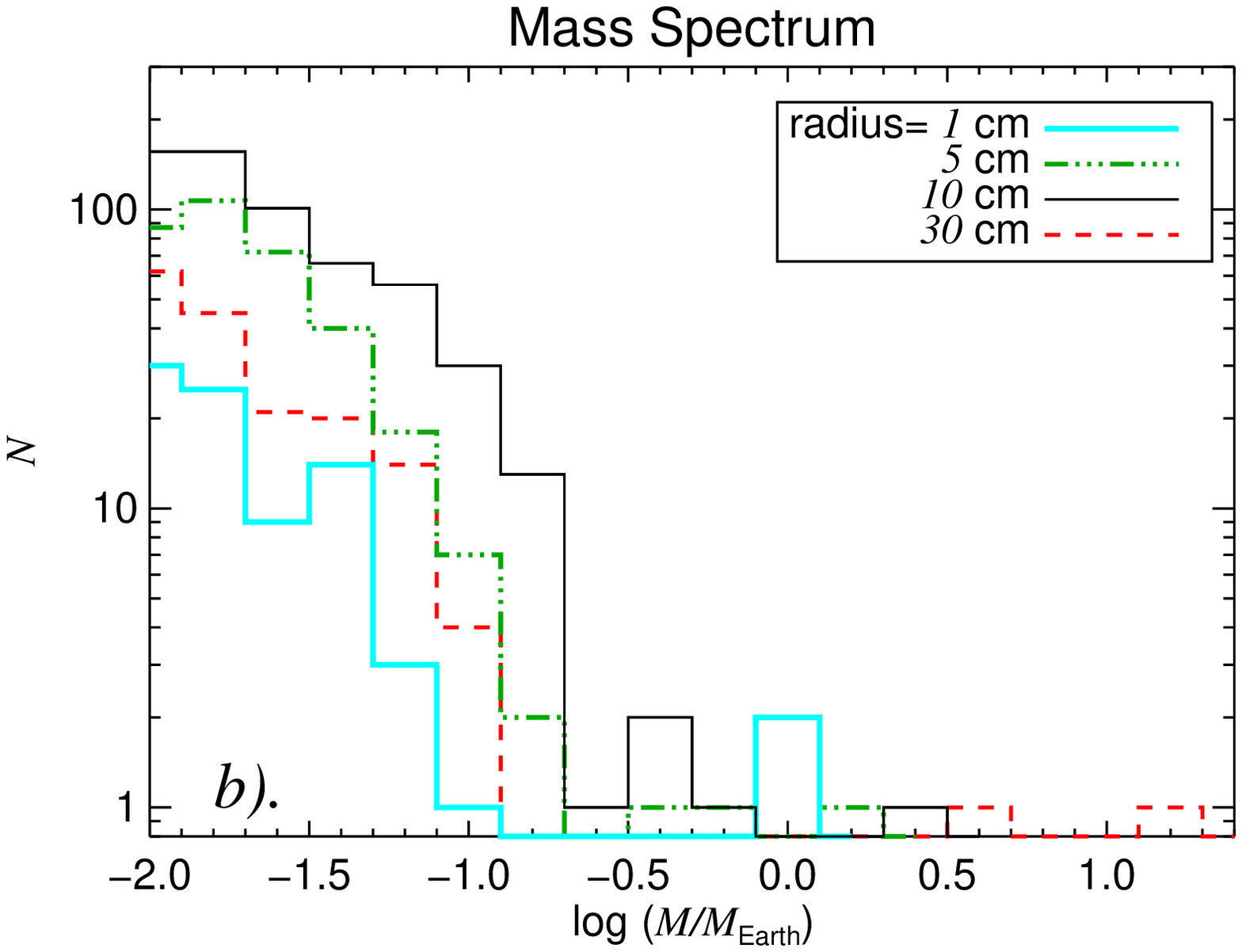}}
  \end{center}
  \caption[]{{\it a).} Time evolution of the maximum concentration of mass for
  different particle radii. From 40\,cm onwards, the drag force is too weak to
  provide enough damping for collapse. For 1\,cm to 30\,cm radii, gravitational
  collapse occurs forming Earth-mass planets. For 5\,cm, a 2\mearthp planet is
  formed. For 10\,cm, the final maximum collapsed mass is around 3\mearth. The
  particles of 30\,cm radius collapse in a planet as massive as 17 Earths. The
  $a_\bullet$=1\,cm are subject to strong coupling and undergo growth on the
  timescale of depletion of gas in the Lagrangian points. This leads to a delay
  in the collapse, taking twice the time it took in the 5-30\,cm case. The
  final mass is 1\mearth.\\ {\it b).} The mass spectrum in the end of
  the simulations.  Along with the super-Earths formed with the
  30\,cm, 10\,cm and 5\,cm particles, dozens of Mars sized and hundreds of Moon
  sized objects were also formed. The two symmetric Trojan Earths in the 1\,cm
  case are apparent.  The runs with particles of $a_\bullet$=40\,cm and
  $a_\bullet$=1\,m were excluded for clarity.}
  \label{fig:timeevo-spectrum}
\end{figure*}

\subsection{Initial conditions}
\label{sect:initial-conditions}

We use a Cartesian box ranging $x,y\in[-2.0,2.0]r_0$, with resolution
256$\times$256. The small extent in radius is justified because we want to
understand what is happening at the vicinity of the planet's orbit at $r_0$ and
the gap it opens. The density profile follows the power law
$\varSigma_g$=$\varSigma_0 r^{-0.5}$ and the sound speed is also set as a power
law $c_s=c_{s_0} r^{-0.5}$.

The gravitational potential is then computed via the Poisson solver and the
initial velocity profile is set to match the condition of centrifugal
equilibrium
\begin{equation}
  \dot\phi^2 = \varOmega_{\rm K}^2 + \frac{1}{r} \left[ \frac{1}{\varSigma_g}\frac{\partial{P}}{\partial{r}} + \frac{\partial{\varPhi_{\rm sg}}}{\partial{r}}     \right] 
  \label{eq:centrifugal}
\end{equation}
The planet is placed initially at ($r$,$\phi$)=($r_0$,0), and the star at
($r$,$\phi$)=(0,$\pi$). To avoid giving the gas and the particles too much
impulse when the planet is introduced in the unperturbed disk, we ramp its mass
up from 0 to its final mass in five orbits, in the way described in de
Val-Borro et al.\ (2006). We computed simulations with companion mass ratios
$q$=$\ttimes{-3}$ (Jupiter) and $q$=$\ttimes{-4}$ (``Neptune''). The quotation
marks are used because calling this mass ratio ``Neptune'' is a jargon, since
the actual mass of the planet is the equivalent to $q$=$\xtimes{5}{-5}$. The
Earth has a mass ratio of $q$=$\xtimes{3}{-6}$.

We use units such that $r_0$=$\varSigma_0$=$GM_\odot$=1. We choose
$c_{s_0}=0.05$ and a Toomre $Q$ parameter of 30 at the position of the planet,
so the gas there is stable against gravitational instability. Assuming that
$r_0$ is the position of Jupiter (5.2 AU) and that $\varSigma_0$=300
g\,cm$^{\rm -2}$, the disk has $\ttimes{-2} M_\sun$ of gas within the modeled
range.  

For the solids, we use $\ttimes{5}$ Lagrangian numerical particles, and
the interstellar solids-to-gas ratio of $\ttimes{-2}$. Each numerical
particle therefore is a super-particle containing $\ttimes{-9} M_\sun \simeq
\xtimes{3}{-2} M_{\rm Moon}$ of material. The super-particle formalism
considers that each numerical particle is an ensemble of a large number of
individual smaller physical particles of radius $a_\bullet$. These particles
share the same position and velocity, interacting gravitationally by their
collective mass (the mass of the super-particle). The aerodynamics, however, is
controlled by the radius $a_\bullet$, which in turn means that there is free
space between the physical particles, so that each of them exposes its whole
surface area to the nebular gas. 

We stress that the mass resolution of solids in the models presented in
this paper is not much greater than that used by Beaug\'e et al.\ (2007). The
main difference between this study and theirs lies in the global
character of our study; the greater number of numerical particles; and
the radius $a_\bullet$ of the individual pebbles and boulders, which
translates into a much stronger drag force. 

We survey several particle radii. The dimensionless friction time as a function
of particle size is found by plugging \Eq{eq:coeff-general} into
\Eq{eq:friction-time}, which yields
\begin{eqnarray}
  T_{_f}&=&\tau_{_f}\varOmega_K \nonumber \\
     &=&\frac{\sqrt{32\pi}}{\Kn'\Ma} \frac{\lambda \rho_\bullet}{\varSigma_g}\frac{(\Kn'+1)^2}{\left(\Kn'^2C_D^{\rm Eps}+C_D^{\rm Stk}\right)}
     \label{eq:dimensionless-friction}
\end{eqnarray}
where we already substituted $\rho_g$=$\varSigma_g/(\sqrt{2\pi}H)$. Here,
$H=c_s/\varOmega_K$ is the pressure scale height. We consider that the
particles have an internal density $\rho_\bullet$=3 g\,cm$^{\rm -3}$. The mean
free path $\lambda$ is 
\begin{equation}
  \lambda=\frac{\mu_{\rm mol}}{\rho_g\sigma_{\rm mol}}
  \label{eq:mean-free-path}
\end{equation}
where $\mu_{\rm mol}=\xtimes{3.9}{-24}$ g is the mean molecular weight of a 5:1
H$_{\rm 2}$-He mixture, and $\sigma_{\rm mol}=\xtimes{2}{-15}$ cm$^{\rm 2}$ is
the cross section of molecular hydrogen. For our densities and sound speed, it
corresponds to 20 cm at the inner radius $r$=0.3, and to 1.3 m at the
outer radius $r$=2.0. 

The result of \Eq{eq:dimensionless-friction} for our choice of parameters (at
the position of Jupiter's orbit) is shown in \Fig{fig:frictiontime2}a
for the grid of Knudsen and Mach numbers. \fig{fig:frictiontime2}b shows
a slice of the grid at the subsonic regime. For particle of 1m diameter, the
coupling due to \Eq{eq:dimensionless-friction} is 50\% looser than
predicted by Epstein law. A factor 2 in the friction time is seen at 2m
diameter between \Eq{eq:dimensionless-friction} and the Stokes law. 

The particles are initialized as to yield a surface density following the same
power law as the gas density, and their velocities are initialized to the
Keplerian value. 

We use reflective boundaries and damp waves in the way described in de
Val-Borro et al.\ (2006). Particles are removed from the simulation if they
cross the inner boundary or if they approach the giant planet by less than 1/5
of its Hill's radius.

\section{Simulations with single particle species}
\label{sect:single-species}

In \Fig{fig:jupiter} we show the time evolution of the disk under the influence
of a $q$=$\ttimes{-3}$ companion, for different particle radii.  Each run has
only one particle size, but as the gas density does not change significantly
between the runs, we just show the gas for the $a_\bullet$=1\,cm case. 

\subsection{Collapse in the Lagrangian points L4 and L5}
\label{sect:lagrangian-collapse}

As the planet opens a gap in the gas, the particles also move out of the
co-rotational region, in the same manner seen in Paardekooper (2007) and
Fouchet et al.\ (2007). The solids at the border of the gap are expelled and
those in the immediate vicinity of the planet are accreted. The particles
inside the co-rotational region librate in horseshoe orbits. The stable leading
(L4) and trailing (L5) Lagrangian points retain high gas densities even
after the planet has carved a deeper gas gap in its orbit, which has a
beneficial effect for the particle concentration. Due to the presence of high
gas densities, the Lagrangian islands are not only a region of convergence of
streamlines, but also a region with higher pressure than its surroundings.  The
drag force therefore forces the particles into them, also damping the motion
caused by eventual perturbations that could otherwise make a particle drift
away from it. These effects combined make L4 and L5 highly stable points in the
motion of a solid particle.

At 20 orbits, an asymmetry is seen in the particle concentration between L4 and
L5, as the trailing Lagrangian point is more efficient in trapping than the
leading one. The 10\,cm, 30\,cm and 1\,m particles achieve high concentrations
in the vicinity of L5, while experiencing depletion in L4.  The 1\,cm particles
are too coupled to the gas to be affected by particle-gas drift. 

At 50 orbits, the concentration in the Lagrangian points has increased by two
orders of magnitude relative to the initial condition in the 10\,cm, 30\,cm and
1\,m runs. The particles of 10\,cm and 1\,m still present an azimuthally
extended cloud of material in L4 and L5, but the particles of 30\,cm radii have
already concentrated into a small swarm spanning but a few grid cells.
Inspection of the snapshot reveals that the maximum mass in this swarm is of
0.03\mearth. The L4 concentration is more extended, but the maximum density is
greater, achieving 0.25\mearth, already exceeding the mass of planet Mars (0.1
\mearth). 

At the end of the simulation at 100 orbits, the swarms of particles in the L4
and L5 points of the $a_\bullet$=1\,m run remain unbound. We ran for additional
50 orbits, but no progress in the maximum mass was seen. If collapse happens,
it requires timescales longer than 150 orbits. The total mass in L4 is
0.29\mearth, peaking at 0.05.  The L5 point has 1.9\mearthp in total, with
maximum mass concentration of 0.3\mearth. The 10\,cm and 30\,cm particles
underwent collapse at the Lagrangian points, with the gravitational
fragmentation being more
efficient for the 10\,cm particles than for the 30\,cm ones. For the
$a_\bullet$=30\,cm case, what appears in \fig{fig:jupiter} as a single clump at
L4 has a mass of 0.18\mearth. The L5 point is still azimuthally extended, with
a total mass of 2.5\mearthp but maximum concentration of only 0.27\mearthp by
the end of the simulation. 

The $a_\bullet$=10\,cm particles underwent collapse in both Lagrangian points,
L5 harboring a 2.6\mearthp planet, L4 a 0.6\mearth.  In
\Fig{fig:timeevo-spectrum}a we plot the time evolution of the maximum mass of
solids for different runs. In addition to the runs showed in \fig{fig:jupiter}
we add runs with particles of 5\,cm and 40\,cm radii.  Collapse in the
Lagrangian points occurs for the 5\,cm case as well, forming a planet of
2\mearth. In this figure, the difference between a run where collapse occurred
and a run where collapse did not occur is readily apparent by the behavior of
the time-series. The non-collapsed ones are very noisy at late times, as the
number of particles in a cell fluctuates up and down. When collapse is
achieved, the maximum mass stays constant unless more mass is accreted. This
gives the time series a ladder-like appearance, as seen in the figure for the
5,10, and 30\,cm cases. Collapse is hindered for $a_\bullet$=40\,cm onwards.

The 1\,cm particles present an interesting behavior. They are so strongly
coupled to the gas that their collapse does not occur at the same time-scale,
as seen from \fig{fig:timeevo-spectrum}a. Instead, as \fig{fig:jupiter}
evidences, it occurs on the timescale of depletion of gas in the tadpole
orbits. As the gap is cleared and its depth increases, the gas clouds in the
Lagrangian islands shrink in size. As the particle are strongly coupled, they
are forced to concentrate as the cloud shrinks, eventually achieving high
densities. As the time series of \Fig{fig:timeevo-spectrum} shows, after
100 orbits the steady increase due to gas clearing gives place to a runaway
growth that lasts for about 20 orbits. In the end, one gravitationally bound
planet encerring one Earth mass of solids - purely out of 1\,cm sized pebbles -,
is formed in each stable Lagrangian point. 

\subsection{Collapse at the gap edge vortices}
\label{sect:vortex-collapse}

Concurrently, at the edges of the gap, the considerable density gradient
resulting from the gap opening process excites the RWI, leading to a
large generation of potential vorticity. At fifty orbits, two vortices are seen
to have been excited by the planet at the outer edge of the gap, seen in
\fig{fig:jupiter} at 5 and 10 o'clock. The effect of these vortices in the
motion of the solids can be readily seen in the $a_\bullet$=1\,cm run, as even
for these tightly coupled particles, the concentration reaches an order of
magnitude higher than in the immediate surroundings. 

In the $a_\bullet$=10\,cm run, as the particles are more loosely coupled to the
gas, the effect of the anti-cyclonic motion is better appreciated.  The
particles are forced in spiral trajectories towards the center of the vortices,
raising the density of solids by another order of magnitude when
compared to the $a_\bullet$=1\,cm particles.

In the $a_\bullet$=30\,cm and $a_\bullet$=1\,m runs, the coupling is too loose
to form the extended structure seen for  the $a_\bullet$=1\,cm and 10\,cm
particles. However, the looseness is a benefit as long as the goal is to
increase the concentration of solids. As the coupling weakens, the particles
are not forced away from the center, and concentrate more efficiently. A
massive clump of particles is seen in the 4 o'clock vortex in the
$a_\bullet$=30\,cm run, that already concentrates 2\mearthp of solid material.
High particle concentration is also seen for the $a_\bullet$=1\,m particles,
but they do not seem to get dense enough to achieve gravitational collapse.
Instead, they form a very azimuthally extended belt of particles at the outer
and inner edge. No collapse is seen at the inner edge of the gap in any of the
runs. At 100 orbits, the vortices have merged into a single giant vortex.
Inside it, in the 30\,cm run, the collapsed mass underwent runaway growth of
solids, reaching 17\mearth. The 10\,cm particles have a maximum mass in the
vortex of 0.3\mearth. The 1\,m particles show a similar maximum mass, of
0.25\mearth. The high mass achieved in the 30\,cm run is quite likely
overestimated, since it is seen that the efficient and unimpeded particle drift
had the effect of feeding this radial region with virtually all particles
present in the simulation. Such a situation may be made quite different in a
more realistic case, where particle drift is stalled by turbulence, for
instance. 

In \fig{fig:timeevo-spectrum}b, we show the mass spectrum at the end of the
different simulations. In addition to the super-Earths, two planets in the
0.5-0.8\mearthp  range were formed out of 5\,cm particles, and other two in the
0.3-0.5 range with the 30\,cm particles. Dozens of Mars-sized planets in the
0.08-0.3\mearthp range, along with hundreds of smaller Moon-sized objects, were
also formed in all simulations. 

\section{A spectrum of particle sizes - segregation and the counter-intuitive
role of self-gravity}
\label{sect:multiple-species}

\begin{figure*}
  \begin{center}
    \resizebox{\textwidth}{!}{\includegraphics{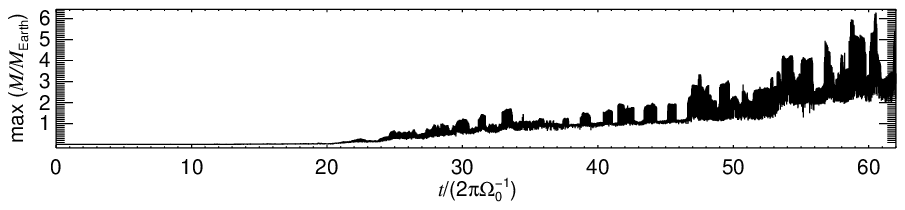}}
    \vspace*{1mm}
    \resizebox{\textwidth}{!}{\includegraphics{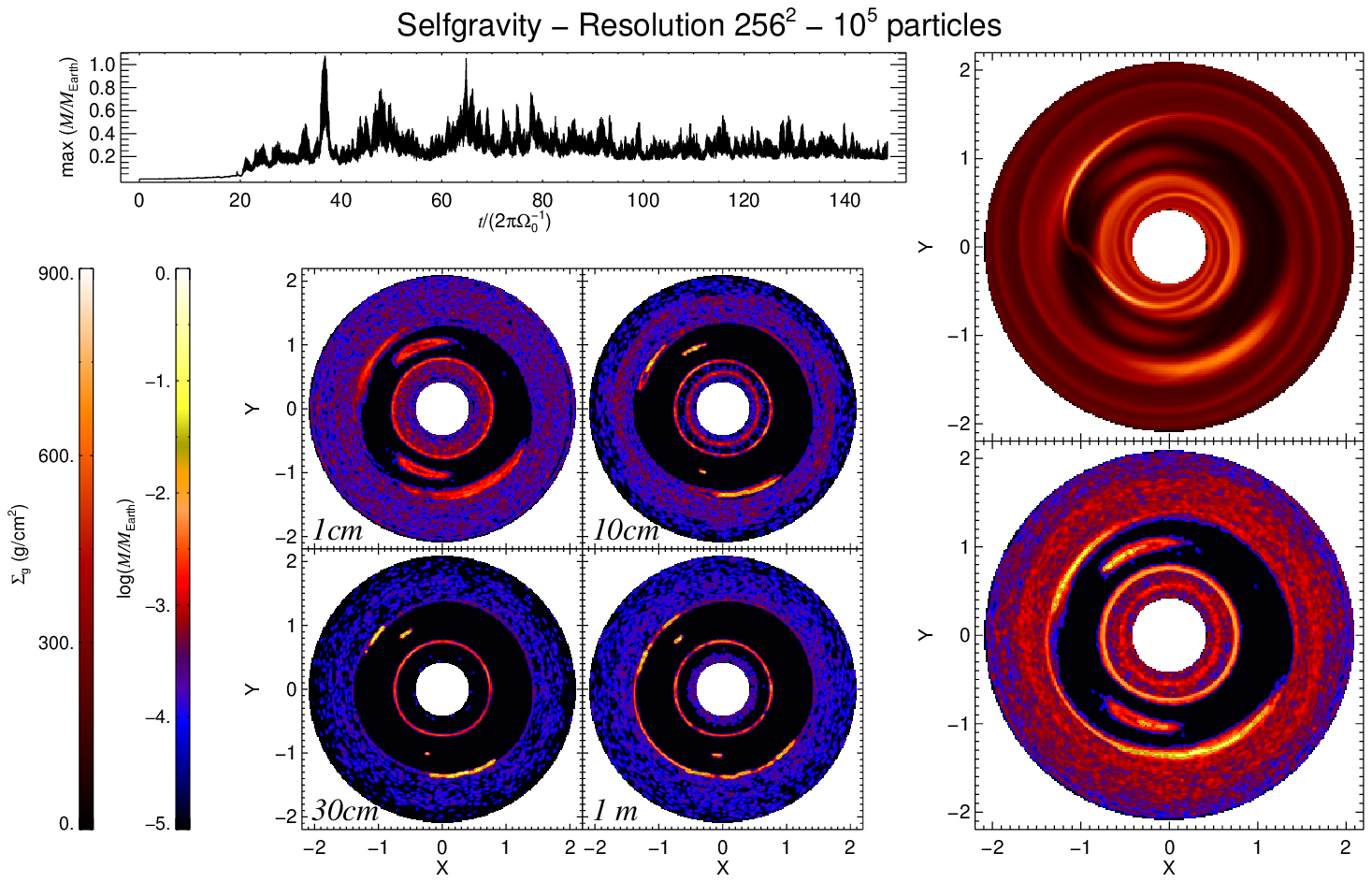}}
  \end{center}
  \caption[]{Comparison between the runs with resolution $256^2$, $10^5$
  particles and multiple particle species, with and without self-gravity (upper
  and lower panels, respectively). Counter-intuitively, self-gravity is seen to
  work against collapse as in the second run the maximum mass is never high
  enough to allow it. In the presence of self-gravity, the tadpole orbits are
  modified, and gas tides from the massive vortices can be disruptive for
  planets forming within them. The motion of the particles inside the vortices
  are also modified in the presence of self-gravity. Notice in particular how
  the 30\,cm and 1\,m particles are spatially split in the vortex at 10
  o'clock, near L5. Collapse proceeds only if the grid resolution is
  refined (see text and \Fig{fig:high-resolution}.)}
  \label{fig:fourspecies}
\end{figure*}

To understand the effect of self-gravity in the runs, we perform a control run
with only gas drag. To diminish the computational time, we include in this run
a spectrum of particles radii, including four species: 5, 10, 30, and 100\,cm.
Each particle species is represented by 1/4 of the total number of particles.
To compare with this non-selfgravitating run, we also compute a
self-gravitating run with a size spectrum. These runs not only shed light onto
the role of self-gravity, but can also discern the possible artificial effects
introduced by the single-species approximation used by us so far.

\subsection{Excluding self-gravity: gas drag alone assembles super-Earths}
\label{sect:exclude-self}

In \fig{fig:fourspecies}a we show the time evolution of the maximum mass in the
non-selfgravitating run, along with the disk appearance in the gas phase as
well as the contribution of each particle species in the solids phase. As
stated before, the drag provides a very efficient damping, so that the
particles of all species except 1\,cm concentrate in the L4 and L5 as well as
in vortices as early as 50 orbits. The 30\,cm and 1\,m particles successfully
concentrate all its remaining particles that lie in the co-orbital region in a
single cell in each of the stable Lagrangian points.  The 30\,cm particles
concentrate with 0.66\mearthp in L5 and 0.04 in L4; the 1\,m particles with
0.54\mearthp in L5 and 0.1 in L4. The concentration of 10\,cm particles is less
efficient, with a maximum concentration of only  25\% of the particles in L5
(which nonetheless means 0.13\mearth). Of the 2.1 Earth masses of material in
L5, the representation is 0.25\mearthp in 1\,cm particles, equal shares of
0.66\mearthp of 10 and 30 \,cm, and 0.54\mearthp  of 1\,m particles. The L4
point has 0.55\mearth, distributed 0.23, 0.17, 0.04, and 0.1\mearthp for 1, 10,
30, and 100\,cm, respectively.

At the outer edge of the gap, as more material is available, the concentration
achieves higher masses. Without self-gravity, the clumps cannot collapse,
quickly dispersing and regrouping instead. The maximum mass then is highly
fluctuating. After 60 orbits, it has grown to 2 Earth masses, but sporadicly
reaching as high as 6\mearth, due to gas drag {\it alone}. The vortex
closest to L5 seen in \fig{fig:fourspecies}, a snapshot at 62 orbits,
concentrates 2.9\mearthp in the densest cell. The 1\,m particles have a
maximum concentration of 2.75\mearth, similarly to the 30\,cm ones, which peak
at 2.3\mearth. It shows that the different particle species preferentially
concentrate in different cells, a result of the different drag they feel. The
same was seen in the Lagrangian points. The 10\,cm is more extended, peaking at
0.1\mearth, a relatively low mass. The leading vortex presents the same
qualitative behavior, with a peaking mass of 1.64\mearth, 10, 30, and 100\,cm
particles showing highest concentration of 0.3, 1.3 and 1.25\mearth,
respectively.

\subsection{Including self-gravity: collapse hampered}
\label{sect:include-self}

When self-gravity is considered (\fig{fig:fourspecies}b), the accretion is seen
to be stalled. Sparse episodes of high particle concentration happen at
$\sim$38 and 65 orbits, reaching maximum masses of 1\mearth, but the collapse
of this mass did not occur and the clump quickly dispersed.  After a hundred
orbits, the maximum mass was still at the 0.2\mearthp level.  The L4 point was
cleared of particles compared to the non-selfgravitating run, displaying
0.7\mearthp of solid material, more than half of it in 1\,cm particles. The
highest concentration is of 0.19\mearth, which is mostly represented by 10\,cm
particles, contributing 0.17\mearth, 100\% of the 10\,cm particles remaining in
the L4 vicinity. The totality of 30\,cm particles in the region are also
concentrated in a single cell, but its mass is of only 0.04\mearth, and
although spatially close to the 0.17\mearthp clump of $a_\bullet$= 10\,cm, they
are not at the same cell. The 1\,m particles still show a slightly extended
cloud, with total mass 0.1\mearth, some degrees away from both 10\,cm and
30\,cm concentrations. The tadpole of particles around L5 is still highly
extended spatially.  We ran the simulation for additional 50 orbits, but the
conditions remained unchanged. In particular, the 3 nearby clumps of different
particle species did not collapse into a single body. 

There are four reasons as to why collapse did not proceed as in the
single species runs. First, the mass of solids was equally split in particles
of different size. The 1\,cm particle retain 1/4 of the mass, and they
concentrate very poorly due to their short friction time. This mass is thus
effectively removed from the mass of potentially collapsable bodies. Running
for longer times to allow the shrinking Lagrangian gas clouds to squeeze the
1\,cm particles into a collapsed body (as occurred in the single species
$a_\bullet$=1\,cm run after 150 orbits) did not produce the same results, as
seen in the time series in the lower panels of \fig{fig:fourspecies}.

Second, the gravitational potential of the massive particles acts to
de-stabilize the Trojan orbits. As the mass in the Lagrangian points grow, the
massless approximation ceases to apply, and the body starts to librate around
the otherwise stable point. As the mass increases, the librations increase in
amplitude and lead the other particles into close encounters with the giant,
that are thence accreted or ejected from the system. In the limiting case that
the mass of the Trojan body becomes comparable to the mass of the planet
itself, the amplitude of libration would become so high that an encounter
between the two would occur. Beaug\'e et al.\ (2007) find that a 0.15\mearthp
object is enough to de-stabilize the orbits of other bodies in the
vicinity of L4. 

The effect of this libration in our simulations is evident when comparing
\fig{fig:fourspecies}a and \fig{fig:fourspecies}b. Instead of concentrating at
L4 and L5 as the massless particles do, the massive particles display an
azimuthally extended structure, evidence of the enhanced librating motion. 

Third, the inclusion of gas gravity leads to tides that can be disruptive
for a prospective planet (Lyra et al.\ 2008b). In a simple yet
informative approximation, the tides can be taken as proportional to the radius
$R$ of the clump and to the gradient of the gravitational acceleration
which, by the Poisson equation, is proportional to the local value of the
density, $F_T \propto R \rho_g$. For a spherical clump of constant
density $\rho_p=3 M_p/(4\pi R^3)$, the self-gravitational pull it exerts on its
own surface is $F_G$=$GM_p/R^2\propto R \rho_p$.  The ratio $F_T/F_G$ is
therefore proportional to the gas-to-solids ratio.  For a protoplanet
forming inside high-pressure regions such as vortices or the Lagrangian clouds,
the gas tides can lead to destruction or significant erosion of the forming
planets (Lyra et al.\ 2008b).

Fourth, a common feature of all simulations is that the particles of
different radii tend to concentrate in different locations within the tadpole
region. This is somewhat similar to the effect of self-gravity. Gas drag taps
energy from the Keplerian motion, so the stability conditions on the Lagrangian
points are modified. As gas drag depends on radius, the location of the stable
points of the 3-body problem with gas drag also depend on particle size. In
other words, the L4 and L5 points of the restricted 3-body problem are defined
as points where there is a balance between the gravitational attraction
between the 2 massive bodies and the centrifugal force. When including
gas drag, a third force comes into play in the particle motion, and the stable
points will be displaced accordingly. In general, a particle of a given size
will librate about its own particular stationary point. Numerical and
analytical investigations by Peale (1993) and Murray (1994) confirm that the
location of the stable points is a function of particle radius. Asymmetries
between L4 and L5 are also expected from the analytical treatment, which are
seen in our simulations as well, with L4 shifting further away than 60\degreep
ahead of the planet, while L5 is displaced closer behind it. In some extreme
cases, the stable points can vanish altogether. As the drag force increases and
L5 approaches the planet, it can merge with the shifted L2 point. L4
experiences the same as it moves further out and merges with the shifted
L3 point.  Both Murray (1994) and Peale (1993) find a limiting location
of 108\degreep ahead of the planet for L4. At this maximum angular separation,
the merging with L3 takes place and the leading stationary point disappears.
For a 13\mearthp proto-Jupiter, Peale (1993) finds that L4 does not exist for
objects smaller than $a_\bullet$=15\,m. L5 is seen to be more stable, but the
stable point of a  $a_\bullet$=50\,cm  particle is expected to lie only a few
degrees behind the proto-Jupiter.  In this location, they speculate, the wake
of the planet (not taken into account in their model) might effectively
eliminate L5. 

Increasing the mass of the perturber to that of Jupiter's present mass
tends to increase stability and to bring L4 and L5 closer to the
``classical'' locations predicted by the restricted 3-body problem. In a gap
homogeneously depleted by 1 order of magnitude relatively to the initial
density, the shift for the 10\,m particles is less than 2\degree. However,
the analysis of Peale (1993) and Murray (1994) did not consider the
presence of higher gas densities in the (classical) Lagrangian points as the
gap is cleared. As we see, it has an effect similar to a potential well,
keeping the particles around the classical tadpole. As the 1\,cm particles have
shorter friction time, the gas trap is more efficient, and the prediction of
their particular L5 getting too close to the planet, or L4 merging with L3 is
avoided as long as a local pressure maximum is present at the classical L4 and
L5. The more loosely coupled 1\,m particles had their L4 shifted to 90\degreep
ahead of the planet, and L5 to 50\degreep behind it. 
\section{Resolution study}
\label{sect:resolution}

\begin{figure*}
  \begin{center}
    \resizebox{\textwidth}{!}{\includegraphics{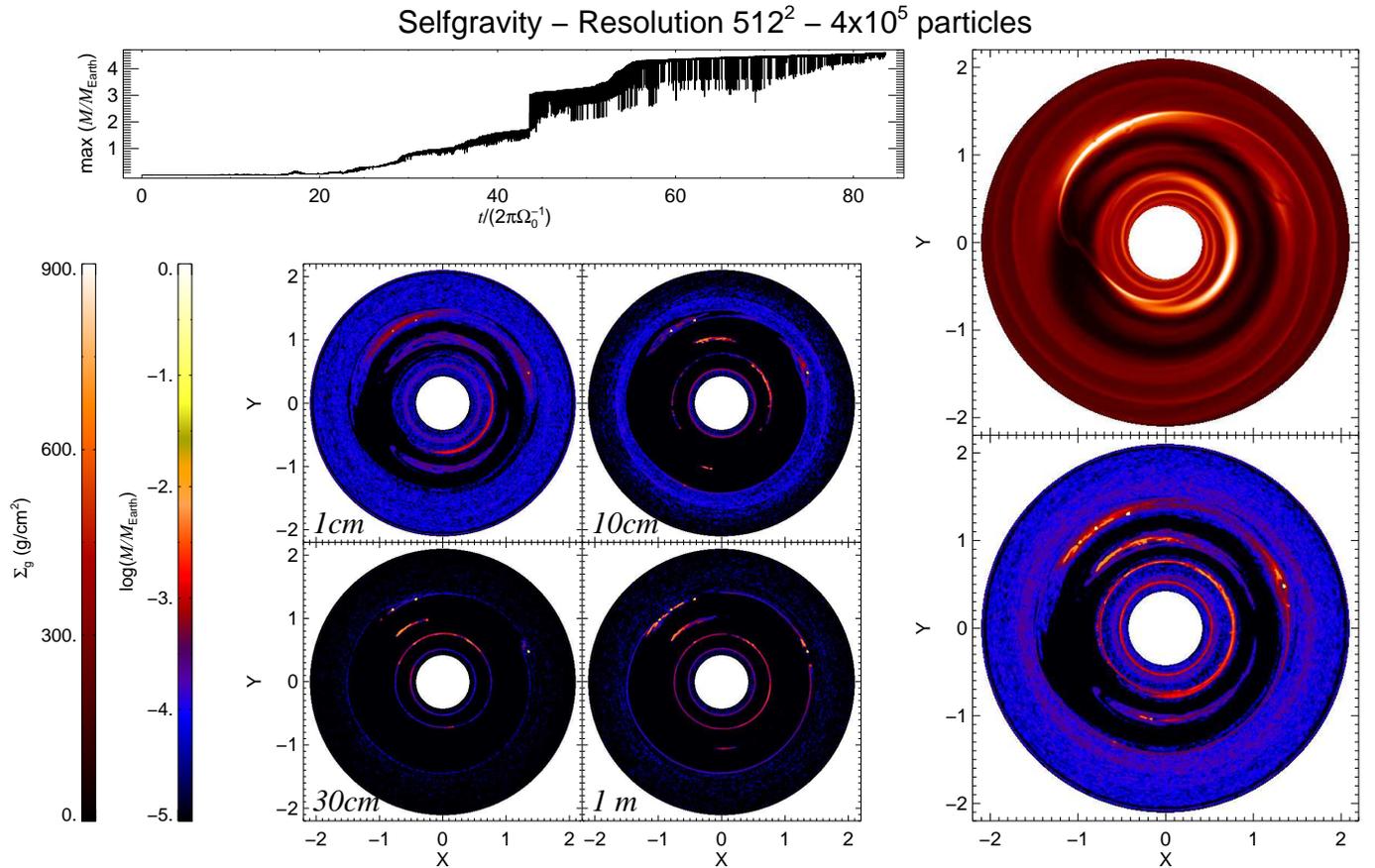}}
  \end{center}
  \caption[]{Results of the high-resolution run ($512^2$ grid points and
  $\xtimes{4}{5}$ particles) with multiple particle species. Four rocky planets
  form at the outer edge of the gap, the most massive one with 4.5\mearth. They
  are easily spotted in the solids plot as very bright small dots. A movie of
  this simulation can be found at {\tt
  http://www.astro.uu.se/{$\sim$}wlyra/planet.html}}
  \label{fig:high-resolution}
\end{figure*}

\begin{figure*}
  \begin{center}
    \resizebox{\hfwidth}{!}{\includegraphics{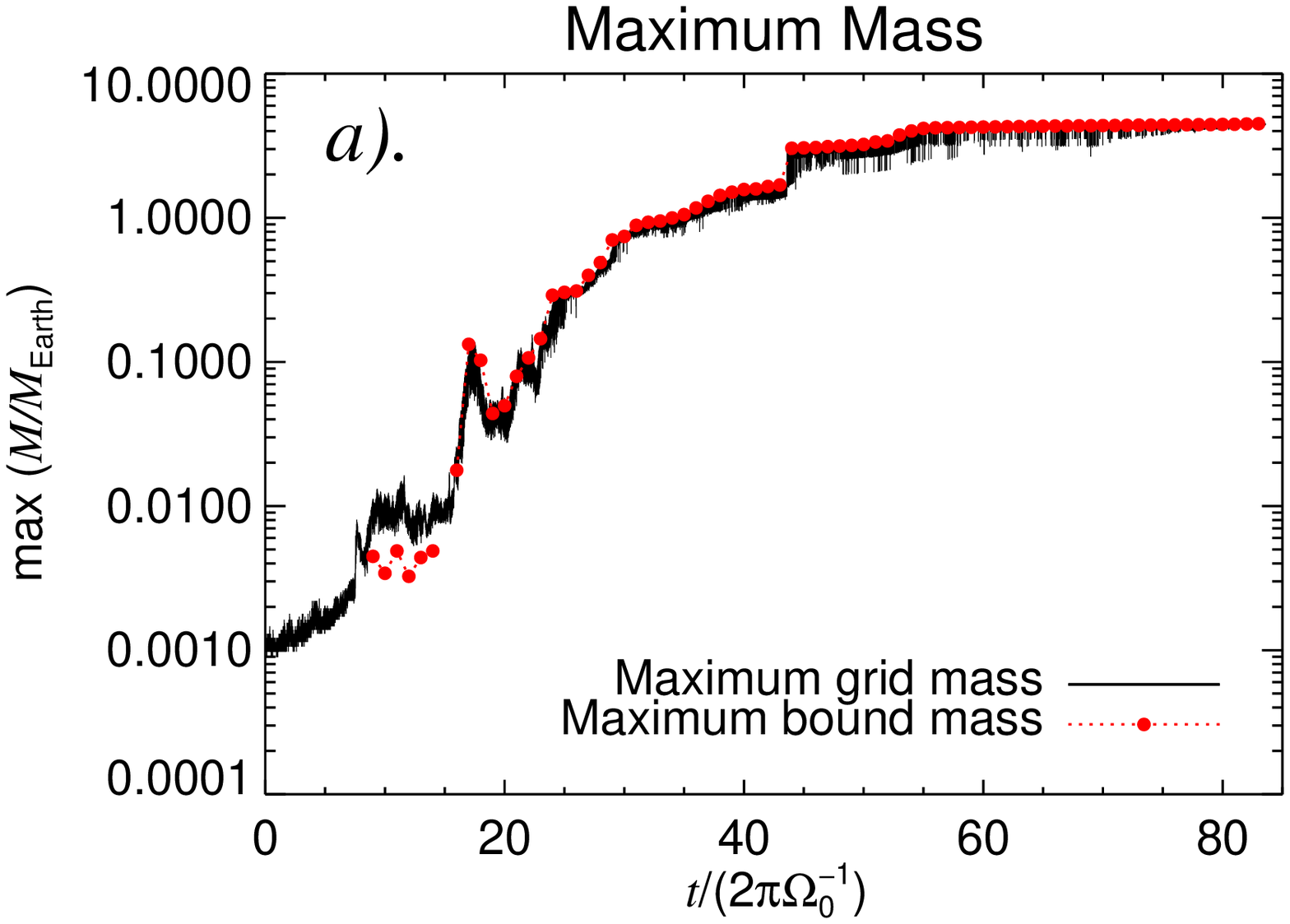}}
    \resizebox{\hfwidth}{!}{\includegraphics{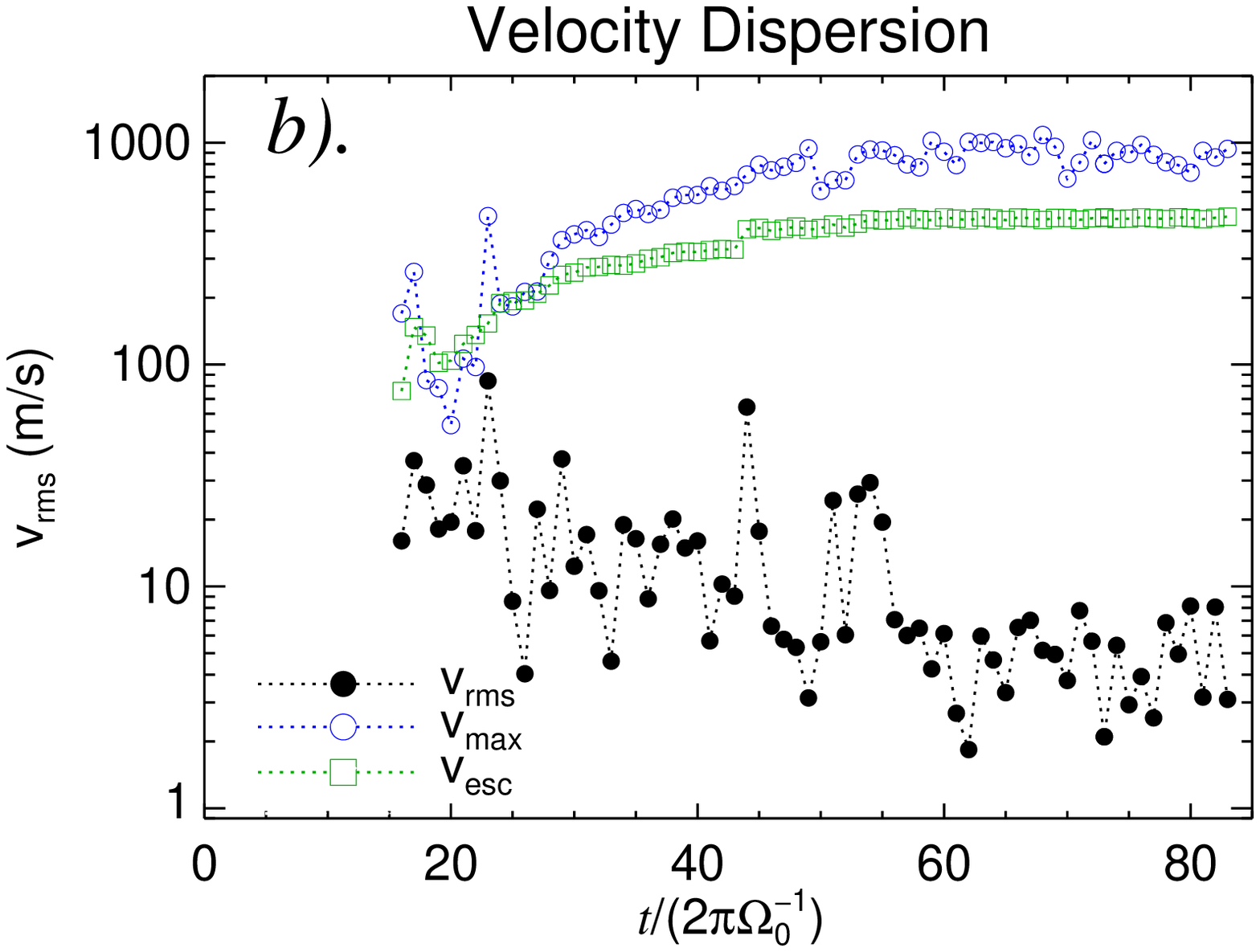}}
  \end{center}
  \caption[]{Time-series of the most massive clump present in
  \fig{fig:high-resolution}.  {\it a).} The maximum mass in a grid cell and the
  maximum bound mass. Even though the Hill's radius exceeds the dimension of a
  grid cell, the planet has most (or all) of its mass within a single cell.
  This is evidence of subgrid compactness.\\ {\it b).} The internal
  velocity dispersion \vrms~ of the planet, compared with its escape velocity
  \vesc~ defined at the Hill's radius. Throughout most of the simulation, \vrms
  is below 10\mps. The maximum internal speed is plotted for comparison.  It
  usually exceeds the escape velocity, so some particles are not bound to the
  planet.} 
  \label{fig:time-series-bound}
\end{figure*}

Motivated by the failure of the run just presented above to assemble massive
gravitationally bound structures, we explore the effects of particle and
grid resolution in our simulations. We first raise the total number of
particles to $N_p$=400\,000, to verify the effect of particle resolution. The
mass of the disk is the same, so the mass of an individual super-particle
decreases, being now $\xtimes{2.5}{-10} M_\odot \simeq
\xtimes{7}{-3}M_{\rm Moon}$. This first run has the same grid resolution as
used before, 256$^{\rm 2}$. The second is twice as fine, 512$^{\rm 2}$.

The $N_{\rm x}\times N_{\rm y}$=256$^{\rm 2}$ and $N_p=\xtimes{4}{5}$ run does
not show major differences when compared to the simulation with same resolution
but only $\ttimes{5}$ particles. The same behavior of sparse episodes of high
concentrations but never achieving critical densities is seen. At the
end of the simulation, the maximum mass is still around only 0.2\mearth. We
conclude from this that changing the particle resolution by at least a factor 4
does not change the results significantly. 

On the other hand, the situation changes considerably when changing the grid
resolution. In the run with $N_{\rm x}\times N_{\rm y}$=512$^{\rm 2}$ and
$N_p$=$\xtimes{4}{5}$ (\fig{fig:high-resolution}), the maximum mass steadily
increases towards 1\mearthp in 30 orbits. Inspection of the snapshots reveals
that this high concentration occurs inside the vortices excited in the outer
gap. At fifty orbits, the leading vortex shows two planets, one of 1.43\mearth,
and a smaller one of 0.38\mearth. Unlike the 256$^{\rm 2}$ run, the mass peaks
of different particles species occur at the same cell, attesting to the
boundness of the structures. The first planet is 57.6\% composed of 30\,cm
particles, 35.0\% of 10\,cm, 6.5\% of 1\,m and 0.9\% of 1\,cm particles. The
second is 87\% composed of 30\,cm particles, about equal shares (6.5\%) of
10\,cm and 1\,m particles, with only trace amounts of 1\,cm particles.  

The trailing vortex also shows two gravitationally bound planets, both of high
mass. The most massive one has 3.1\mearth, its composition of 1, 10, 30, and
100\,cm particles being 0.2\%, 17.9\%, 63.0\%, and 18.9\%, respectively.  The
other planet is of 1.9\mearth, being constituted by 0.2\%, 27.8\%, 48.9\%, and
23.1\% of 1, 10, 30, and 100\,cm, respectively. 

A common trait of these planets is, therefore, that they are formed by a
majority of 30\,cm particles, with approximately equal shares of 10\,cm and
1\,m particles. This is expected, since for our choice of parameters, the
30\,cm particles are those for which the drift due to gas drag is maximum. The
1\,cm are too well coupled to the gas to contribute significantly to the growth
of terrestrial planets inside the vortices. For reasons of load imbalance, we
terminated the simulation at 83 orbits, when a large fraction of the
computational time was idle and one orbit took 6 hours in 64 processors.
The most massive planet had grown to 4.5 \mearthp by then. The other 
planets formed at the outer edge of the gap show masses of 4.36, 4.14, and 0.80 
Earth masses.

In \fig{fig:time-series-bound}a we show the time evolution of the mass of this
massive planet. The black solid line represents the maximum mass of solids
contained in a single grid cell. The red dashed line marks the maximum mass
that is gravitationally bound. We decide for boundness based on two criteria.
First we consider the clump defined by the black line, and calculate the center
of mass of its particles. The Hill's sphere associated with this mass is drawn,
centered on the center of mass. As the Hill's sphere encompasses
more/less than a grid cell, particles inside/outside are added/removed from the
total mass, and the center of mass and Hill's radius recomputed. The process is
iterated until convergence. After the clumps' mass and Hill's radius are
defined, we compare the internal velocity dispersion \vrms~ of its constituent
particles with the escape velocity of the enclosed mass, defined at the Hill's
radius. If \vrms $<$ \vesc, we consider that the cluster of particles is
gravitationally bound. As seen in \fig{fig:time-series-bound}b, the internal
velocities are usually lower than 10\mps. We also plot the maximum speed and
escape velocity of the planet (defined at the Hill's radius). The maximum speed
is usually greater than the escape velocity, which means that not all particles
present in the cluster are actually bound, and the planet (as we define it) can
lose mass during the accretion process. However, the low \vrms~ compared to
\vesc~ attests that the vast majority of the particles is gravitationally
bound. 

At the end of the simulation, the gas in co-rotation is still spread over the
whole horseshoe region, so a massive loss of particle from L4 is observed.  The
same process was seen in the other runs, with single and/or multiple species.
But in this case, the effect is more severe as the L4 point of the 30\,cm
particles disappeared. At the end of the simulation, a small cloud of
2 $M_{\rm Mars}$ of 10\,cm particles is observed in the tadpole region 
around L4, peaking at a maximum mass of 3.5 $M_{\rm Moon}$. L5 presents 
3.3\mearthp of solid material, but still in extended clouds. The boundness
analysis shows that these clouds are fragmented into $\approx$20 sub-Mars sized
bodies of mass between 1-5 lunar masses.



\section{Neptune-mass perturber}
\label{sect:neptune}

\begin{figure*}
  \begin{center}
    \resizebox{\textwidth}{!}{\includegraphics{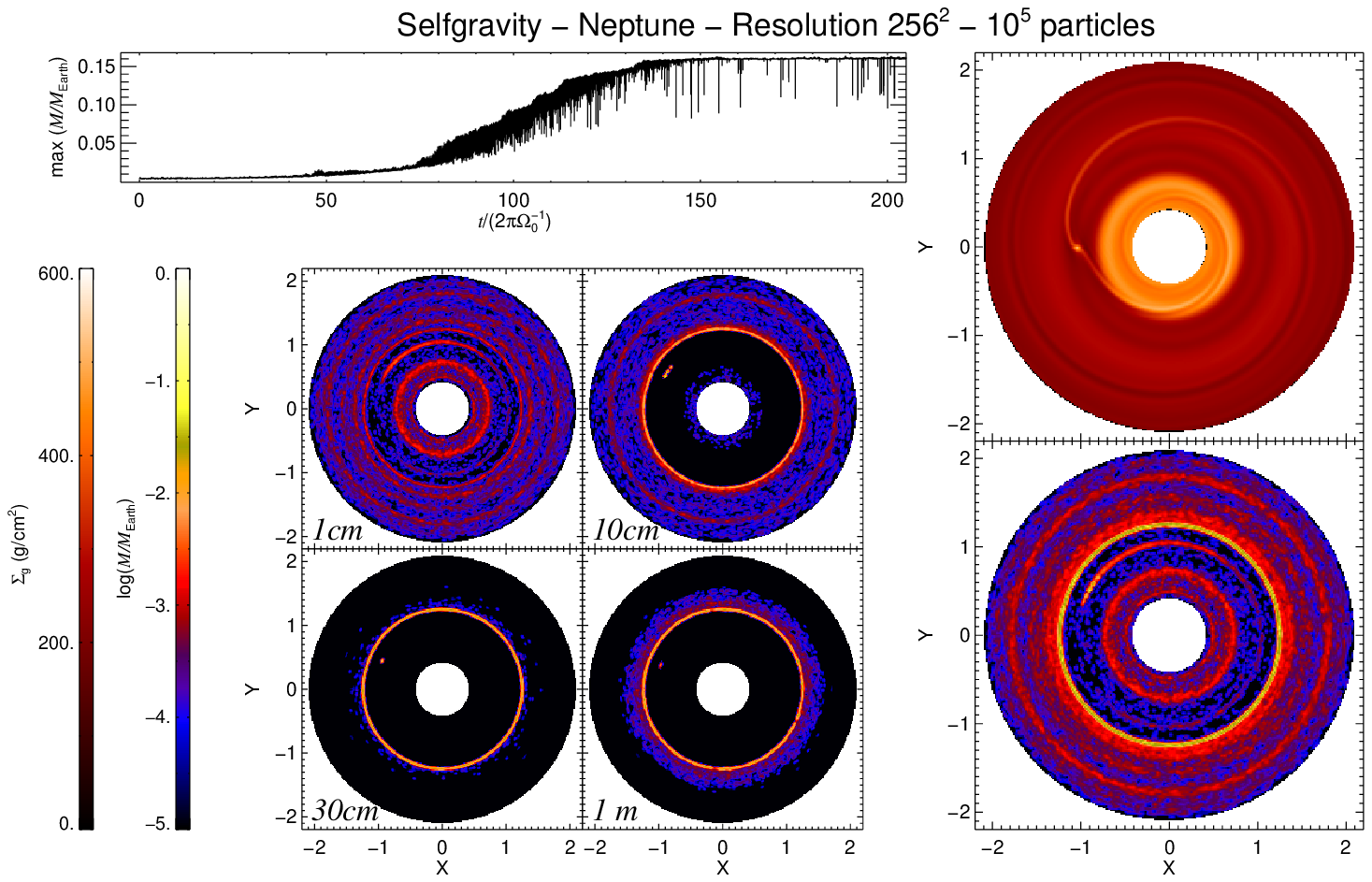}}
  \end{center}
  \caption[]{Same as \fig{fig:high-resolution} but for a Neptune-mass
  perturber. The Lagrangian point L4 has vanished and the L5 shifted to a
  position much nearer to the planet than in the Jupiter case. The wake of the
  planet does not destroy the stability of the shifted L5, and a Trojan of 1.6
  $M_{\rm Mars}$ was formed.} 
  \label{fig:neptune}
\end{figure*}

In this section, we consider the case of a giant planet perturber of
mass ratio $q$=$\ttimes{-4}$, dubbed ``Neptune''. This case is important to assess
since, according to our current understanding, a forming gas giant is expected
to spend a long time (of the order of millions of years) with a mass similar to
this value - corresponding to the phase II of the model of Pollack et al.\
(1996). Even models that predict a faster transition from Neptune to Jupiter
mass (Klahr \& Bodenheimer, 2006) still predict timescales of ($\sim$10$^{\rm
5}$ years). Therefore, when the perturber has achieved Jupiter's mass, the state
of the solids subdisk should be more similar to the state left by a
Neptune-mass perturber than to the unperturbed disk of particles we have used
so far.

We observe that when the perturber has a smaller mass, a more
pronounced asymmetry between the L4 and L5 point is observed, as expected from
the analysis of Peale (1993) and Murray (1994). The 1\,m and 30\,cm
sized particles experience more depletion, with their L4 point having
vanished altogether and the L5 shifted to but a few degrees behind the planet
(\Fig{fig:neptune}). The 10\,cm particles also
experience depletion but not as severe as the larger
particles. The 1\,cm particles are well coupled and remain in co-rotation as no
deep gas gap is carved. 

The shifted L5 points of the 30\,cm and 1\,m particles concentrate about only 0.01
\mearthp of solids, each. Nevertheless, a Trojan planet of 0.16\mearthp
was formed at the vicinity of L5, its bulk consisting of 99.4\% of particles of
10\,cm radii. A second bound clump of 0.09\mearth, also consists of a large
majority of 10\,cm particles, is observed at the vicinity of L5, 0.39\,AU away
from the former. 

We conclude that a Neptune Trojan can only be formed with a very narrow range
of particle species around 10\,cm, at least for our choice of parameters.
A simulation at high-resolution (with 512$^{\rm 2}$ grid points and
$\xtimes{4}{5}$ particles) showed the same behavior for the first 100 orbits.

A distinct difference from the Neptune runs when compared to the Jupiter
runs is that there are no visible vortices formed at the edge of the gas gap,
even when running as long as 200 orbits. This was unexpected, since the gap is
shallow when compared to the one carved by the Jovian tides, but deep enough to
excite the RWI. Therefore, there is no clear reason as to why vortices do not
form. The solution was hinted by de Val-Borro et al.\ (2007), who notice the
same feature. They identify it as being due to the Cartesian grid, as vortices
are seen in a cylindrical run. Furthermore, in de Val-Borro et al.\
(2006), where several codes were compared in the specific problem of a planet
opening a gas gap, vortices are seen in some of the inviscid runs with
cylindrical codes. Indeed, we ran simulations with the cylindrical version of
Pencil, and some weak vortices were excited after 100 orbits. This is
readily understandable in view of the fact that for a flow with cylindrical
symmetry, a Cartesian grid has exaggerated numerical dissipation for the same
resolution ($r\Delta{\phi}$=$\Delta{y}$). To make matters worse, the azimuthal
modes responsible for the RWI, are more coarsely resolved in a Cartesian grid.
We are drawn to the conclusion that the combination of both drawbacks quenched
the growth of the unstable modes of the RWI in the case of the shallow Neptune
gap. 

In the cylindrical run at two hundred orbits, the vortices had
trapped large amounts of particles, with a few cells achieving masses above
0.1\mearth. However, the cylindrical Poisson solver- which relies on
discretization of the analytical potential based on continuous Hankel
transforms (Toomre 1963, Binney \& Tremaine, 1987) - does not ensure that
a particle is free of self-acceleration. Therefore we do not trust its
accuracy to draw definitive conclusions on the possibility or impossibility of
gravitational collapse in the cylindrical runs.

We stress that the expulsion of particle of radii $>$10\,cm from the
co-rotational region during the Neptune-phase does not imply that these
particles will not be present when the giant planet achieves Jupiter's mass. As
the planet grows in mass, the width of the gas gap increases. This has the
positive effect of feeding the co-rotational region with fresh larger
particles from the outer and inner edge of the narrow and shallow gap carved
during phase II. Moreover, it is reasonable to suspect that growth by
coagulation  should be continuously replenishing the population of these
particles, as the pebbles sweep up dust grains that remain in the co-rotational
region.

\begin{figure}
  \begin{center}
    \resizebox{\hfwidthsingle}{!}{\includegraphics{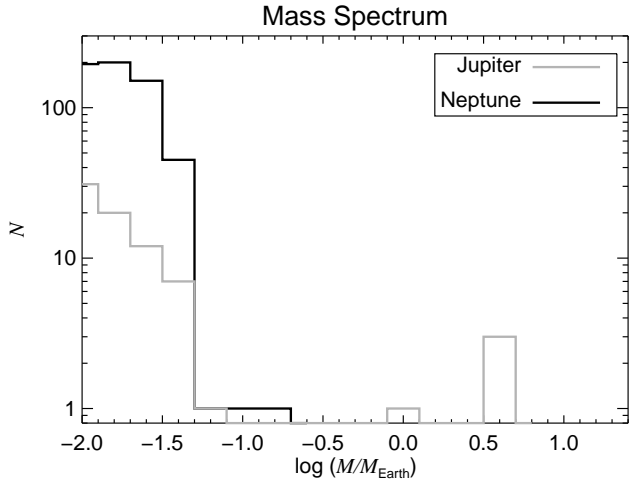}}
  \end{center}
  \caption[]{Mass spectrum at the end of the Jupiter (Sect.~\ref{sect:resolution}) and
  Neptune (Sect.~\ref{sect:neptune}) simulations. The most massive planet in the Neptune case is
  the Trojan of 0.16 \mearth. It is followed by another Trojan of 0.09\mearth.
  In the Jupiter case the most massive planets are formed within the vortices
  at the outer edge of the gas gap. It includes the three super-Earths and a 8
  $M_{\rm Mars}$ planet. The Jupiter simulation was terminated at 83 orbits. 
  At that time, the mass in the tadpole orbits was split into 
  19 lunar-sized objects in the mass range 1-5 $M_{\rm Moon}$. Other 26 
  gravitationally bound objects of mass between 0.5 and 1 $M_{\rm Moon}$ are 
  also observed in the co-rotational region.}
  \label{fig:nepjupspec}
\end{figure}

We show in \fig{fig:nepjupspec} the mass spectrum at the end of the
Neptune simulation, comparing it with the one from the Jupiter case 
(Sect.~\ref{sect:resolution}). In addition to the two Trojans, the Neptune 
run also shows a smaller planet, of mass 4.6 times that of the Moon, which 
was formed at the outer edge of the gap. The outer edge also displays hundreds 
of other Moon-sized objects. In the Jupiter case the three super-Earths 
are conspicuous in the plot. The smaller 0.80 \mearthp planet is 
also visible. Of the seven lunar-sized bodies in the bin centered at 
$\log(M/$\mearth$)$=$-1.4$ ($M$$\approx$$4 M_{\rm Moon}$), three are in the 
co-rotational region. Their masses are 4.8, 4.3, and 4.2 $M_{\rm Moon}$. 
Other sixteen lunar-sized bodies in the mass range 1-4$M_{\rm Moon}$ are 
also observed in the co-rotational region. As more mass is trapped in the 
bigger planets, the Jupiter run shows a smaller number of Moon-mass 
gravitationally bound clumps when compared to the Neptune case.

\section{Summary and conclusions}
\label{sect:conclusions}

We have undertaken simulations of low mass self-gravitating disks with gas and
solids. While the gas is gravitationally stable ($Q \approx 30$), the solid
phase undergoes rapid collapse in the Lagrangian points of a giant planet. A
companion with the mass of Jupiter (mass ratio $q$=$\ttimes{-3}$) produces
Earth-mass Trojan planets for particle radii up to $a_\bullet$=30\,cm.  The
particles of $a_\bullet$=40\,cm and 1\,m remained unbound. The 10\,cm
and 30\,cm particles underwent collapse at the Lagrangian points, with the
gravitational fragmentation being more efficient for the 10\,cm
particles than for the 30\,cm
ones. In the $a_\bullet$=10\,cm case, the particles underwent collapse in both
Lagrangian points, L5 harboring a 2.6\mearthp planet, L4 a 0.6\mearth. The
30\,cm particles show only a low mass 0.1\mearthp at L4, and an extended
unbound swarm at L5. Particles of 5\,cm radius assembled in Trojans of 1.8
\mearth, 0.8 and 0.5 \mearth. The 1\,cm particles present an interesting
behavior.  As they are too well coupled to the gas, their density increase
primarily not due to their mutual attraction, but due to the shrinking of the
gas cloud retained in the tadpole region. Their collapse therefore occurs on
the timescale of gas depletion in the L4 and L5 points. Two symmetric Trojans
of 1\mearthp are formed out of particles of $a_\bullet$=1\,cm after 150 orbits.
The boundness of the formed planets is confirmed as the internal velocities
are much lower than the escape velocity.

Fast rocky planet formation also occurs in the vortices the giant
planets induce at the edges of the gas gaps they open. In this case, the
30\,cm particles set the record of highest concentration, by collapsing into a
super-Earth encerring as much as 17 Earth masses. The mass is likely to be
overestimated, since the vortex captured virtually all of the influx of
particles from the outer disk, but this result nonetheless illustrates that the
efficiency of vortex trapping for particles this size is superb. For other
particle radii, the mass spectrum shows that dozens of Mars-sized planets were
formed, along with hundreds of Moon-sized objects.

We compare runs with single and multiple particle species, finding that gas
drag modifies the streamlines in the tadpole region around the classical
L4 and L5 points. As a result, particles of different species have their
stable points shifted to different locations. This brings down the mass of the
Trojan planets, as now the clumps are segregated spatially by size, each of
them having less mass available for assemblage. As a result, collapse is
hindered in a low-resolution run with 256$^{\rm 2}$ grid points and
$\ttimes{5}$ particles equally distributed in mass and number among four
species (1, 10, 30, and 100 cm). Counter-intuitively, a run with the same
parameters but without self-gravity achieved higher mass concentrations (up to
6\mearth).  We conclude that the gravity of the solids modifies the
stability of the tadpole orbits. Inside the massive vortices, the tidal forces
from the gas also stall the gravitational growth of the solids into planets.
The same negative results are observed when the number of numerical
super-particles is raised by a factor 4. 

Collapse resumed when the grid resolution was refined by a factor 2, producing
3 super-Earth mass planets at the outer edge of the gap. The
most massive one has 4.5\mearthp by the end of the simulation. The other 
super-Earths are of 4.36 and 4.14 \mearth. In addition, a fourth, smaller, 
planet of 8.0 $M_{\rm Mars}$ was also formed within the gap edge vortices. These planets 
are composed primarily 
of 30\,cm particles ( $\approx$50\%), with smaller and almost equal shares of
10\,cm and 1\,m, and only trace amounts of 1\,cm particles. Judging by
their mass and location, these objects may be the embryos that gave rise to
planet Saturn. Although the 
distance of formation of Saturn in this model seems too close to Jupiter, it is
not at all unlikely that Saturn was indeed formed in this orbital position. The
ice giants Uranus and Neptune are presently located in regions of the solar
system where the dynamical timescales are too large and the densities are too
low to account for their current masses (Thommes et al.\ 2002, and references
therein). This is an indication that they were formed further in and,
therefore, that the giant planets displayed a much more compact spacing in the
early Solar system than they present today. Our results seem to corroborate
this scenario

When the mass of the perturber is reduced to that of Neptune, the
asymmetry between L4 and L5 is accentuated.  The L5 point of the particles of
$a_\bullet$=10\,cm moves to $\approx$35\degree behind Neptune, and the
$a_\bullet$=1\,m to $\approx$25\degree. The L4 point was shifted too far ahead
of the planet and eventually lost all particles, a behavior attributed to its
merging with the shifted unstable L3 point (Peale 1993, Murray 1994).  Of the
particles retained at L5, the ones of 10\,cm concentrated into a 1.6 $M_{\rm
Mars}$ Trojan planet. 

One question to ask is if the formation of Trojan bodies as massive as
terrestrial planets is so easily achievable, why we do not see it in the Solar
System. The answer might lie in the fact that, according to recent models by
Morbidelli et al.\ (2005), all Trojan orbits of the Jovian system were
de-stabilized when Jupiter and Saturn crossed the 2:1 mean motion resonance.
The initial Trojan population of Jupiter was lost and a new one was captured.
Without the gas to damp their motions and increase the number density, the new
Trojan population could not assemble into
rocky planets. This scenario raises the possibility that in extrasolar
planetary systems with only one giant or with giants that did not undergo the
destructive resonance crossing that Jupiter and Saturn underwent, Trojan 
Earth-mass companions to the giant planets are common. This includes the giants in
Earth-like orbits in a list of potentially habitable stellar systems. 

Of course, it might as well be that the formation scenario we present is overly
simplistic and that some important piece of physics that prohibits the process
is missing. We did not include, for instance, the possibility of destructive
collisions between boulders. Checking the velocity dispersion at the bound
clumps, we find that they are typically  lower than 10\mps for a formed planet.
As the initial stages of collapse, however, the speeds are greater, 10-30\mps,
eventually reaching as fast as 80\mps. These speeds are comparable to or
larger than the break-up collisional speeds ($\sim$10\mps, Benz 2000).
These high collision speeds indicate that collisional fragmentation will
play an important role during the gravitational collapse in a more realistic
coagulation-fragmentation model (Brauer et al.\ 2008a). On the other hand,
the fact that collapse occurs for particles of 1-10\,cm radius is particularly
relevant since they are too small to be easily destroyed by collisions.
Moreover, the escape velocities of the formed clumps are high enough so that
most debris of catastrophic collisions might remain bound. Johansen et al.\
(2008) find that cm-sized fragments of such collisions are easily swept up away
from the midplane by turbulent motions. This leaking is anticipated to not
occur in the cases presented in this paper, where planets are formed inside
vortices. As vortices do not have vertical shear and revolve at the Keplerian
orbital rate (Klahr \& Bodenheimer 2006) the sedimenation of the solids layer
does not trigger the Kelvin-Helmholtz instability (Johansen et al.\ 2006b) when
this sedimentation happens inside a vortex. The sedimentation is therefore more
efficient, which helps collapse.

Our neglecting of coagulation is also an issue that causes pause.  Solid
bodies grow by sweeping up smaller dust grains, so coagulation raises the
possibility that the trapped rocks and boulders might breach the meter-size
barrier inside the gap edge vortices and Lagrangian gas clouds. If so, they
would produce km-sized bodies that are too loosely coupled to undergo
gravitational collapse in the way presented in this paper. Brauer et al.\ (2008b)
has indeed showed that growth to kilometer-size is highly favored within gas
pressure maxima. However, the timescale for coagulation seems to be slow
($\sim$1000 yr) compared to the timescales we observe for gravitational
collapse in all cases except for the formation of Trojan planets with the
$a_\bullet$=1\,cm particles. In this case, the timescales are comparable and we
can expect coagulation to influence the growth. In particular, coagulation onto
the 1\,cm particles can aid on replenishing the population of 10\,cm and 30\,cm
particles lost during the Neptune phase.

Once a cluster of particles collapses to form a single object, aerodynamical
drag ceases to be the most important driver of particle dynamics. Instead the
planet enters the regime of gravitational drag in which it interacts with its
own gravitational wakes. Since we solve for both the particle gravity (that
causes the wakes) and gas gravity (that makes the wakes backreact on the
particles), our simulations in principle resolve this stage as well, although
limited by the grid resolution. However, the drag influence of the planet on
the gas is strongly exaggerated, since the influence of particles is always
spread over the nearest three grid points in each direction. The friction time
is also still that of the individual rocks, where as a solidified body of a few
thousand kilometers in size should have a much longer friction time. A better
treatment would thus be to replace the ensemble of particles by a single
particle representing the planet. This would also allow a much longer
integration time, and we plan to go this way to model the long term evolution
of the planet system in a future project.

An immediate question to ask is how (or if) the collapse would occur in three
dimensions. Johansen \& Klahr (2005), Fromang \& Papaloizou (2006) and
Lyra et al.\ (2008a) show that the particles are stirred up by the hydromagnetic
turbulence to form a layer of finite vertical thickness, maintained by
turbulent diffusion. We performed a 3D simulation of planet-disk interaction in
spherical coordinates, similar to those of Bate et al.\ (2003), Kley et al.\
(2005) and Edgar \& Quillen (2008), but inviscid instead of viscous. The
Lagrangian points of the planet do not change much in 3D, with the scale height
being about the same as in the unperturbed disk case. Fromang et al.\ (2004) and
Lodato (2008) calculate the effects of self-gravity in the vertical extent of
the disk, showing that the thickness is reduced by the disk's self-gravity.
This flattening of the scale height in self-gravitating disks bring it closer
to the 2D configuration. 

Of course, we are only assessing this by simple estimates based on isolated
bits of physics done by individual works. A definite answer to this question
has to be addressed by a 3D simulation that combines these effects.  

The collapse of the solids is triggered by the gravitational influence of a
perturber, but more fundamentally due to the presence of long-lived, 
high-pressure regions: the vortices and the accumulation of gas in the Lagrangian
points. As such, a giant is not necessary for the rapid formation of rocky
planets. Paardekooper et al.\ (2008) show that passing binaries can stir the
material in the disk. Such encounters usually last for long times, and
therefore gravitational collapse of the boulders might happen in such case.
Vortices similar to the ones presented in this paper, excited by a giant
planet, are also expected at the border of the dead zone (Varni\`{e}re \&
Tagger, 2006; Lyra et al.\ 2008b). Therefore, the accumulation into
rocky planets shown to occurs inside the vortices induced by 
a giant planet should also happen inside these dead zone vortices. 
If so, this paper provides not only a plausible mechanism for the 
formation of Trojan planets and Saturn, but also of the very first 
planetary embryos that - in the core accretion scenario - gave 
rise to Jupiter.

\begin{acknowledgements}
Simulations were performed at the PIA cluster of the Max-Planck-Institut
f{\"u}r Astronomie and on the Uppsala Multidisciplinary Center for Advanced
Computational Science (UPPMAX). This research has been supported in part
by the Deutsche Forschungsgemeinschaft DFG through grant DFG Forschergruppe 759
``The Formation of Planets: The Critical First Growth Phase''.
\end{acknowledgements}

\end{document}